\documentclass[aps,prb,twocolumn,reprint,superscriptaddress]{revtex4-1}
\usepackage{graphicx,color}
\usepackage{amsmath,amssymb,amsthm,bm,braket,ascmac,bbm,dsfont,comment}

\newcommand{\dd}{\mathrm{d}}
\newcommand{\ee}{\mathrm{e}}
\newcommand{\ii}{\mathrm{i}}

\theoremstyle{definition}

\newcommand{\hl}[1]{\textcolor{black}{#1}}
\newcommand{\hlr}[1]{\textcolor{black}{#1}}
\newcommand{\hlm}[1]{\textcolor{black}{#1}}

\newcommand{\bfig}{\begin{figure}\begin{center}}
\newcommand{\efig}{\end{center}\end{figure}}

\newcommand{\putfigraw}[1]{\includegraphics[width=\columnwidth]{#1}}

\newcommand{\psec}{\,\text{ps}}
\newcommand{\K}{\,\text{K}}

\newcommand{\kB}{k_\text{B}}

\newcommand{\hfirst}{\hat{H}_\text{I}}
\newcommand{\hsecond}{\hat{H}_\text{II}}
\newcommand{\hfirstmat}{H_\text{I}}
\newcommand{\hsecondmat}{H_\text{II}}
\newcommand{\hxx}{\hat{H}_\text{XX}}
\newcommand{\hxy}{\hat{H}_\text{XY}}

\newcommand{\hstag}{\hat{H}_\text{stag}}
\newcommand{\hext}{\hat{H}_\text{ext}}

\newcommand{\hextZ}{\hext^\text{Z}}

\newcommand{\hextMS}{\hext^\text{MS}}

\newcommand{\hextmat}{H_\text{ext}}
\newcommand{\hextZmat}{\hextmat^\text{Z}}
\newcommand{\hextMSmat}{\hextmat^\text{MS}}

\newcommand{\hextZnew}{\hat{H}_\text{ext,II}^\text{Z}}

\newcommand{\gapI}{\Delta_\text{I}}
\newcommand{\gapII}{\Delta_\text{II}	}

\newcommand{\pix}{\hat{U}^x_\pi}
\newcommand{\pixd}{\hat{U}^{x\dag}_\pi}

\newcommand{\tfwhm}{t_\text{FWHM}}
\newcommand{\gs}{\ket{\psi_\text{gs}}}
\newcommand{\tini}{t_\text{ini}}

\newcommand{\hS}{\hat{S}}
\newcommand{\hSvec}{\hat{\bm{S}}}

\newcommand{\hP}{\hat{P}}
\newcommand{\hJ}{\hat{J}_\text{spin}}
\newcommand{\hM}{\hat{M}}

\newcommand{\Jspin}{J_\text{spin}}

\newcommand{\dP}{\varDelta P}

\newcommand{\stagXY}{J_\text{stag}}
\newcommand{\stagZ}{H_\text{stag}}

\newcommand{\etauZ}{\eta^\text{u}_\text{Z}}
\newcommand{\etasZ}{\eta^\text{s}_\text{Z}}

\newcommand{\etauMS}{\eta^\text{u}_\text{MS}}
\newcommand{\etasMS}{\eta^\text{s}_\text{MS}}

\newcommand{\tfig}{g_u}
\newcommand{\aniso}{\epsilon}

\newcommand{\gsII}{\ket{\Psi_\text{II}}}
\newcommand{\gsIIp}{\ket{\Psi_\text{II}'}}

\newcommand{\cre}{\hat{c}^\dag}
\newcommand{\ann}{\hat{c}}

\newcommand{\spinorI}{\phi_{\text{I},k}}

\newcommand{\epI}{\epsilon_\text{I}(k)}
\newcommand{\epII}{\epsilon_\text{II}(k)}

\begin{document}
%##############################
%##############################
%##############################
\title{High-harmonic generation by electric polarization, spin current, and magnetization}
\author{Tatsuhiko N. Ikeda}
\affiliation{Institute for Solid State Physics, University of Tokyo, Kashiwa, Chiba 277-8581, Japan}
\author{Masahiro Sato}
\affiliation{Department of Physics, Ibaraki University, Mito, Ibaraki, 310-8512, Japan}
\date{\today}

%##############################
%##############################
%##############################
\begin{abstract}
High-harmonic generation (HHG), a typical nonlinear optical effect, 
has been actively studied in electron systems such as semiconductors and superconductors. As a natural extension, we theoretically study HHG from electric polarization, spin current and magnetization 
in magnetic insulators under
terahertz (THz) or gigahertz (GHz) electromagnetic waves. 
We use simple one-dimensional spin chain models with or without 
multiferroic coupling between spins and the electric polarization,
and study the dynamics of the spin chain coupled to an external ac electric or magnetic field.
We map spin chains to two-band fermions and invoke an analogy of semiconductors and superconductors.
With a quantum master equation and Lindblad approximation, we compute the time evolution
of the electric polarization, spin current, and magnetization, showing that they exhibit clear harmonic peaks.
We also show that the even-order HHG by magnetization dynamics
can be controlled by static magnetic fields in a wide class of magnetic insulators.
We propose experimental setups to observe these HHG,
and estimate the required strength of the ac electric field $E_0$ for detection
as $E_0\sim100$\,kV/cm--1\,MV/cm, which corresponds to the magnetic field $B_0\sim0.1$\,T--1\,T.
The estimated strength would be relevant also for experimental realizations of other theoretically-proposed nonlinear optical effects in magnetic insulators
such as Floquet engineering of magnets.
\end{abstract}
\maketitle

%##############################
%##############################
%##############################
%##############################
\section{Introduction}\label{sec:intro}
Ultrafast nonlinear phenomena in condensed matter systems
have recently attracted much attention
owing to the development of laser science and technology~\cite{Calegari2016}.
A remarkable example is the high-harmonic generation (HHG)~\cite{Brabec2000} in semiconductors~\cite{Ghimire2011}.
A key to this success is that strong mid-infrared laser fields have been available in recent years~\cite{Ghimire2014}.
Since the photon energy $\hbar\Omega$ of the input laser is much smaller than the band gap,
nonlinear dynamics is relevant and HHG is clearly seen~\cite{Schubert2014,Hohenleutner2015,Luu2015,Ndabashimiye2016,Vampa2017,Kaneshima2018}.
Recent solid-state-HHG studies in semiconductors~\cite{Faisal1989,Holthaus1994,Golde2006,Golde2008,Korbman2013,Vampa2015,Wu2015,Ikemachi2017,Ikeda2018a,Xia2019,Navarrete2019}
have been extended, for example, to
Dirac systems~\cite{Mikhailov2008,Yoshikawa2017,Hafez2018,Cheng2019}, superconductors~\cite{Matsunaga2014,Kawakami2018,Yonemitsu2018}, charge-density-wave materials~\cite{Nag2018,Ikeda2018b}, Mott insulators~\cite{Murakami2018,Murakami2018a,Imai2019}, topological insulators~\cite{Bauer2018,Jurss2019}, and amorphous solids~\cite{You2017,Jurgens2019,Chinzei2019}.

%#######
In view of the rapid development of the HHG in electronic systems,
it is natural to ask if it can be realized in spin systems (magnetic insulators) without electronic transitions.
%### laser-spin interaction development
The interplay between light and magnets 
has been intensively studied from the viewpoints of spintronics, magnonics,
magneto-optic effects, Floquet engineering, and so on. 
The study of visible and infrared lasers has a long history and 
the ultrafast spin dynamics driven by such high-frequency lasers has been long studied~\cite{Kirilyuk2010}. 
On the other hand, thanks to the recent development of terahertz (THz) laser science~\cite{Hirori2011,Dhillon2017,Liu2017,Mukai2016}, 
magnetic dynamics driven by THz waves has been explored as well in the last decade. 
Since the photon energy in THz or gigahertz (GHz) range is comparable with those of magnetic excitations in magnetic insulators, 
such low-frequency lasers or electromagnetic waves make it possible to directly create and control magnetic excitations or states. 
Therefore, THz or GHz wave is necessary for mimicking HHG in semiconductors with spin systems. 
In fact, recently, the second harmonic generation originated from magnetic excitations in an antiferromagnetic insulator 
has been observed with an intense THz laser~\cite{Lu2017}. 
In addition to this, various experimental studies of THz-wave driven magnetic phenomena have been done: 
intense THz-laser driven magnetic resonance in an antiferromagnet~\cite{Yamaguchi2010,Mukai2016}, 
magnon resonances in multiferroic magnets 
with the electric field of THz wave or laser~\cite{Pimenov2006,Takahashi2011,Kubacka2014}, 
spin control by THz-laser driven electron transitions~\cite{Baierl2016}, 
dichroisms driven by THz vortex beams in a ferrimagnet~\cite{Sirenko2019}, etc.   
These experimental studies have stimulated theorists in many fields of condensed-matter physics, and as a result, 
several ultrafast magnetic phenomena driven by THz or GHz waves have been proposed and predicted: 
THz-wave driven inverse Faraday effect~\cite{Takayoshi2014a,Takayoshi2014b}, 
Floquet engineering of magnetic states such as chirality ordered states~\cite{Sato2016} 
and a spin liquid state~\cite{Sato2014}, applications of topological light waves to 
magnetism~\cite{Fujita2017a,Fujita2017b,Fujita2018,Fujita2019}, 
control of exchange couplings in Mott insulators with low-frequency pulses~\cite{Mentink2015,Takasan2019}, 
optical control of spin chirality in multiferroic materials~\cite{Mochizuki2010}, 
and rectification of dc spin currents in magnetic insulators with THz or GHz waves~\cite{Ishizuka2019a,Ishizuka2019b}. 
Very recently, Takayoshi et al.~\cite{Takayoshi2019} numerically calculate the HHG spectra in quantum spin models, 
assuming that the applied THz laser is extremely strong beyond the current technique.

%#### Problem
Despite of these activities, it is still difficult to realize a sufficiently strong
laser-spin coupling in the THz and GHz regimes.
A main reason for the difficulty is that the field amplitude of THz laser pulse is quite limited (at most the order of 1\,MV/cm) 
compared with the visible and the mid-infrared lasers. 
In addition, electromagnetism tells us that the light-spin coupling is generally smaller than the light-charge one
roughly by a factor of $c^{-1}$ with $c$ being the speed of light.
Therefore, to find practical experimental ways of HHG in spin systems, 
it is important to study how easy or difficult it is to observe the HHG with a moderate field strength 
at the frequency as large as the spin gap.
The long study of magnetic resonances shows that, when THz or GHz wave at adjusted frequency $\Omega$
is applied to magnetic insulators, we can usually obtain a clear linear response signal.
Thus the question is whether significant signals at $n\Omega$ ($n=2,3,\dots$)
appear as the nonlinear response to THz fields with moderate strength.

% Experimental viewpoint
From the experimental viewpoint, the HHG may be one of the simplest phenomena in nonlinear optical effects 
in magnetic insulators. Therefore, estimating required strength of THz or GHz waves for the HHG contributes 
not only to deepen the understanding of the HHG itself, but also to give us a reference value of required ac fields 
for the realization of the other nonlinear phenomena such as optical control of magnetism~\cite{Mochizuki2010,Fujita2017a,Fujita2017b}, Floquet engineering of magnets~\cite{Takayoshi2014a,Takayoshi2014b,Sato2016,Sato2014}, 
and spin current rectification~\cite{Ishizuka2019a,Ishizuka2019b}.

%#### In this paper 
In this paper,
we theoretically study 
harmonic generation and harmonic spin currents
in magnetic insulators
with ac electric and magnetic fields
within the reach of current technology.
We take account of relaxation
of magnetic excitations
and investigate their dynamics by means of
a quantum master equation,
showing that clear harmonic signals are
present at reasonable field strengths.
We show that the photon energy $\hbar\Omega$
of the driving field does not need to be much smaller
than the spin gap if the relaxation is relevant
as is typically the case with magnetic insulators.
This finding implies that a wider class of
THz and GHz electromagnetic waves are useful
to experimentally observe nonlinear optical
effects in magnetic insulators.

%#####
The rest of this paper is organized as follows.
In Sec.~\ref{sec:setup1},
we introduce an inversion-asymmetric spin chain
as a simple realistic model for 
multiferroic or standard magnetic insulators
in order to study harmonic responses
of electric polarization and spin current.
We transform this model into that of fermions
with two energy bands \hlm{like semiconductors},
and formulate our quantum master equation,
which describes dynamics in the presence of relaxation.
On the basis of this formulation,
we investigate the dynamics and its Fourier spectra
of electric polarization and spin current
in Secs.~\ref{sec:pol} and \ref{sec:sc}, respectively.
Thanks to the introduction of relaxation into
the master equation, those spectra are well-defined
without artificial treatment such as window functions.
We show that harmonic generation and harmonic spin currents
can be produced and detectable by using currently available
lasers.
In Sec.~\ref{sec:mag},
we introduce an anisotropic model including the transverse-field Ising model
in order to study harmonic generation through
nonlinear magnetization dynamics.
This model can be mapped to a fermionic BCS-type Hamiltonian
of superconductors,
and our quantum master equation is also applicable.
We thereby conduct a parallel analysis,
showing that harmonic generation is possible
through magnetization.
We also show that the SHG can be controlled by static magnetic fields.
In Sec.~\ref{sec:experiments},
we discuss how to experimentally detect the harmonic generation
and the harmonic spin currents.
\hl{
In the above sections,
we mainly focus on relatively low-order harmonics ($n=2,3,4,5$),
which could be observable with the currently available
laser strength even though the laser-spin coupling is weak in principle.
In Sec.~\ref{sec:extreme},
we study the harmonic spectra under hypothetical strong fields
of theoretical interest.
We thereby discuss the correspondence between the spin system
and semiconductors or superconductors
at the level of the harmonic spectra.
}
Finally, in Sec.~\ref{sec:conclusions},
we summarize our results
and make concluding remarks.

%##############################
%##############################
%##############################
%##############################
\section{Setup for electric polarization and spin current}\label{sec:setup1}
%#######
\subsection{Model and Observables}\label{sec:model1}
To study nonlinear dynamics of electric polarization and spin current,
we consider a simple spin model in one dimension.
The Hamiltonian is given by
\begin{align}\label{eq:Hfirst}
    \hfirst = \hxx + \hstag,
\end{align}
with
\begin{align}
    \hxx &= J \sum_{j=1}^{2L} (\hS_j^x \hS_{j+1}^x + \hS_j^y \hS_{j+1}^y),\label{eq:Hxx}\\
    \hstag &= \sum_{j=1}^{2L} (-1)^j \left[  \stagXY (\hS_j^x \hS_{j+1}^x+ \hS_j^y \hS_{j+1}^y)+\stagZ  \hS_j^z\right],
\end{align}
where $\hS_j^\alpha=\sigma_j^\alpha/2$ ($\alpha=x,y$, and $z$)
are the spin operators at site $j$
with $\sigma_j^\alpha$ being the Pauli matrices.
Here $\hxx$ represents the isotropic XY model with exchange coupling $J$,
and $\hstag$ consists of the staggered exchange coupling $\stagXY$ and the staggered Zeeman coupling $\stagZ$. 
The number of sites is $2L$, and the periodic boundary condition is imposed.

The Hamiltonian~\eqref{eq:Hfirst} is a simple but
realistic model for one-dimensional quantum magnets.
In fact, the staggered exchange coupling $\stagXY$ often appears in spin Peierls magnets 
such as $\rm CuGeO_3$~\cite{Kiryukhin1996,Arai1996,Palme1996,Haldane1982,Takayoshi2010,Sato2012} and TTF-CA~\cite{Torrance1981,Nagaosa1986,Katsura2009}. The staggered field term 
$\stagZ$~\cite{Oshikawa1997,Affleck1999,Sato2004} is known to exist 
in a class of two-sublattice spin chain compounds such as Cu-benzoate~\cite{Dender1997}, 
$\rm [PMCu(NO_3)_2(H_2O)_2]_n$ (PM=pyrimidine)~\cite{Feyerherm2000},
$\rm KCuGaF_6$~\cite{Morisaki2007,Umegaki2009},
$\rm KCuMoO_4OH$~\cite{Nawa2017},
and $\rm Yb_4As_3$~\cite{Oshikawa1999}. 

The symmetries of the Hamiltonian~\eqref{eq:Hfirst} are as follows.
Both the bond-center and site-center inversion symmetries are broken 
if both $\stagXY$ and $\stagZ$ are nonzero. 
These inversion-symmetry-breaking terms are very important to consider the HHG spectra 
because
even-order HHG signals ($n\Omega$ with $n=2,4,6,\cdots$) generally disappear in inversion-symmetric systems~\cite{Franken1961}.
Note that the Hamiltonian~\eqref{eq:Hfirst} has the global U(1) symmetry around the $S^z$-axis
and the total magnetization $\sum_j \hS_j^z$ is conserved.
In Sec.~\ref{sec:mag}, we switch to another Hamiltonian,
for which it is not conserved,
and discuss the magnetization dynamics and harmonic generation.

%########
We describe the laser-spin coupling by either of the following two effects.
The first one is the Zeeman coupling to the laser magnetic field $B(t)$
along the $S^z$ direction, $\hextZ(t) = -B(t) \sum_{j}[\etauZ +(-1)^j \etasZ]  \hS_j^z.$
Here $\etauZ=g\mu_\text{B}$, $g$ is the $g$ factor,
$\mu_\text{B}$ the Bohr magneton,
and we set $\hbar=1$ throughout this paper.
We assume that $\etasZ\neq0$ when we consider $\hfirst$
because the magnetic field acting on site $j$ is modified in general
by the inner magnetic field.
The part of $-\etauZ B(t) \sum_{j} \hS_j^z$ causes no physical effect
because the total magnetization is also conserved in the presence of $B(t)$.
Thus, when we consider $\hfirst$, we will use
\begin{align}\label{eq:HextZ}
\hextZ(t) = -b(t) \sum_{j}(-1)^j\hS_j^z
\end{align}
with $b(t)\equiv B(t)\etasZ$.

%#########
The second laser-spin coupling is the so-called magnetostriction effect~\cite{Tokura2014}.
This is the coupling of the laser electric field $E(t)$
to the spin-dependent electric polarization proportional to $\hSvec_j\cdot \hSvec_{j+1}$.
We write the coupling term as
$-E(t)\sum_j [\etauMS+(-1)^j \etasMS ]  (\hS^x_j \hS^x_{j+1}+\hS^y_j \hS^y_{j+1})$.
We have assumed that the dot product $\hSvec_j\cdot \hSvec_{j+1}$ is dominated by $\hS^x_j \hS^x_{j+1}+\hS^y_j \hS^y_{j+1}$
correspondingly to our Hamiltonian~\eqref{eq:Hfirst}.
Here $\etauMS$ is a constant converting the spin dot product into the polarization,
and $\etasMS$ is its staggered counterpart.
We note that $\etasMS$ is larger than $\etauMS$ in typical multiferroic materials~\cite{Pimenov2006,Takahashi2011,Kubacka2014,Tokura2014},
and thus neglect $\etauMS$ for simplicity.
The coupling Hamiltonian is therefore given, in this work, by
\begin{align}\label{eq:HextMS}
	\hextMS(t)=-e(t)\sum_j (-1)^j(\hS^x_j \hS^x_{j+1}+\hS^y_j \hS^y_{j+1})
\end{align}
with $e(t)\equiv E(t) \etasMS$.

%#######
The electric polarization
\begin{align}
	\hP = \etasMS \sum_j (-1)^j(\hS^x_j \hS^x_{j+1}+\hS^y_j \hS^y_{j+1})\label{eq:pol}
\end{align}
is the first observable of interest. 
When its expectation value evolves in time as $P(t)=\langle \hP\rangle_t$,
it becomes the source of electromagnetic radiation,
which is useful for the experimental detection.
The radiation power at frequency $\omega$ is given by
\begin{align}\label{eq:powP}
	I_P(\omega) \propto |\omega^2 P(\omega)|^2,
\end{align}
where $P(\omega)$ is the Fourier transform of $P(t)$. 
Before the application of laser, the expectation value of the polarization 
$P_0=\langle \hat P\rangle_{\tini}$ is generally nonzero. 
Since a constant shift of $P(t)$ does not change $I_P(\omega)$,
we will also use $\varDelta P(t)=P(t)-P_0$.
In the following, we show that $I_P(\omega)$ exhibits several peaks
at integer multiples of the driving frequency,
which correspond to the HHG.

\hlm{
We remark that the even-order HHG vanishes
for $\stagXY=0$, at which the system Hamiltonian $\hfirst$
is invariant whereas the polarization $\hP$ is odd
under the (site-center) inversion.
This is exactly the same selection rule for the ``conventional'' HHG in inversion-symmetric semiconductors.
The selection rule is understood in the perturbation regime as follows.
For instance, the second harmonic derives from $P(2\Omega)=\chi^{(2)}E_\Omega E_\Omega$, where $\chi^{(2)}$ is the nonlinear susceptibility
and $E_\Omega$ is the Fourier component of the input field
(see a textbook~\cite{Boydbook} for more rigorous discussions).
By applying the inversion, we also have $-P(2\Omega)=\chi^{(2)}E_\Omega E_\Omega$, where we have used the invariance of $\chi^{(2)}$ in inversion-symmetric systems.
The above two equations imply that $\chi^{(2)}$ and hence $P(2\Omega)$
vanish in inversion-symmetric systems.
Similar arguments hold true for all the even-order HHGs,
which thus vanish in inversion-symmetric systems.
Note that such constraints are not obtained for the odd-order HHGs
that in fact exist both inversion-symmetric and -asymmetric systems.
}

%We note that the polarization~\eqref{eq:pol} is odd under the site-center inversion.
%As a consequence, the even-order HHG vanishes
%if $\stagXY=0$ and the site-center inversion symmetry is retained
%as we will see below.

%###
The spin current is the second observable of interest.
This is defined through the continuity equation for $\hS^z_j$,
and its definition depends on the coupling term.
When we consider the total Hamiltonian $\hfirst+\hextZ(t)$,
the spin current operator is given by
\begin{align}
	\hJ &= \sum_j [J+(-1)^j \stagXY ] (\hS^x_j \hS^y_{j+1}-\hS^y_j \hS^x_{j+1})\label{eq:sc1}.
\end{align}
On the other hand,
when we consider the total Hamiltonian $\hfirst+\hextMS(t)$,
it is given by
\begin{align}
	\hJ &= \sum_j \left\{J+(-1)^j [\stagXY -e(t) ] \right\}
	 (\hS^x_j \hS^y_{j+1}-\hS^y_j \hS^x_{j+1}).\label{eq:sc2}
\end{align}
Both Eqs.~\eqref{eq:sc1} and \eqref{eq:sc2}
are odd under the inversion like the electric current
in semiconductors,
\hlm{and the even-order harmonics vanish when the system is inversion-symmetric (the above argument on the polarization applies equally).}
We will see, in Sec.~\ref{sec:fermion},
the spin currents in our setup are analogous to the charge
currents in semiconductors.

Whereas the electric polarization is measured as the radiation from it,
the spin current is usually measured through conversion to an electric current.
Thus, in discussing the spin current, we use 
the Fourier component $\Jspin(\omega)$ by itself,
rather than the radiation power such as Eq.~\eqref{eq:powP}.

%##############################
%##############################
%##############################
%##########
\subsection{Fermionization}\label{sec:fermion}
Our spin model can be mapped to noninteracting spinless fermions
by means of the Jordan-Wigner transformation~\cite{Sachdev2011}:
$\hS^+_j=\prod_{i(<j)}(1-2\cre_i\ann_i)\ann_j$,
$\hS^-_j=\prod_{i(<j)}(1-2\cre_i\ann_i)\cre_j$,
and $\hS^z_j = 1/2- \cre_j \ann_j$
with $\hS^\pm_j=(\hS^x_j\pm \ii \hS^y_j)/2$.
The Hamiltonian~\eqref{eq:Hfirst} is simplified
by introducing the following Fourier transformations for the odd and even sites:
\begin{align}
\hat{a}_k \equiv \frac{1}{\sqrt{L}} \sum_{j=1}^L \ee^{-\ii k (2j)} \ann_{2j};
\hat{b}_k  \equiv \frac{1}{\sqrt{L}} \sum_{j=1}^L \ee^{-\ii k (2j+1)} \ann_{2j+1}\label{eq:fourier_b}
\end{align}
with $k=\pi m/L$ $(m=0,1,\dots,L-1)$.
By introducing the two-component fermion operator
$\spinorI\equiv{}^\text{t}(\hat{a}_k,\hat{b}_k)$,
one obtains
\begin{align}
	\hfirst &= \sum_k \spinorI^\dag \hfirstmat(k) \spinorI,
\label{eq:H1}
\end{align}
where $\hfirstmat(k)$ is a $2\times2$ matrix representation
in our basis and given by
\begin{equation}
	\hfirstmat(k) =  J\cos k \sigma_x -\stagXY \sin k \sigma_y -\stagZ \sigma_z.
\end{equation}
The two eigenvalues of $\hfirstmat(k)$ are $\pm\epI$
with $\epI=\sqrt{(J\cos k)^2+(\stagXY\sin k)^2+\stagZ^2}$,
which define the two energy bands.
Thus the band gap, i.e. spin gap, is given by
\begin{align}
	\gapI = 2\sqrt{\stagXY^2 +\stagZ^2},
\end{align}
%### 
where 
$\stagXY$ and $\stagZ$ are assumed to be smaller enough than $J$.
The band structures are illustrated in Fig.~\ref{fig:band}.

%########
\bfig
\putfigraw{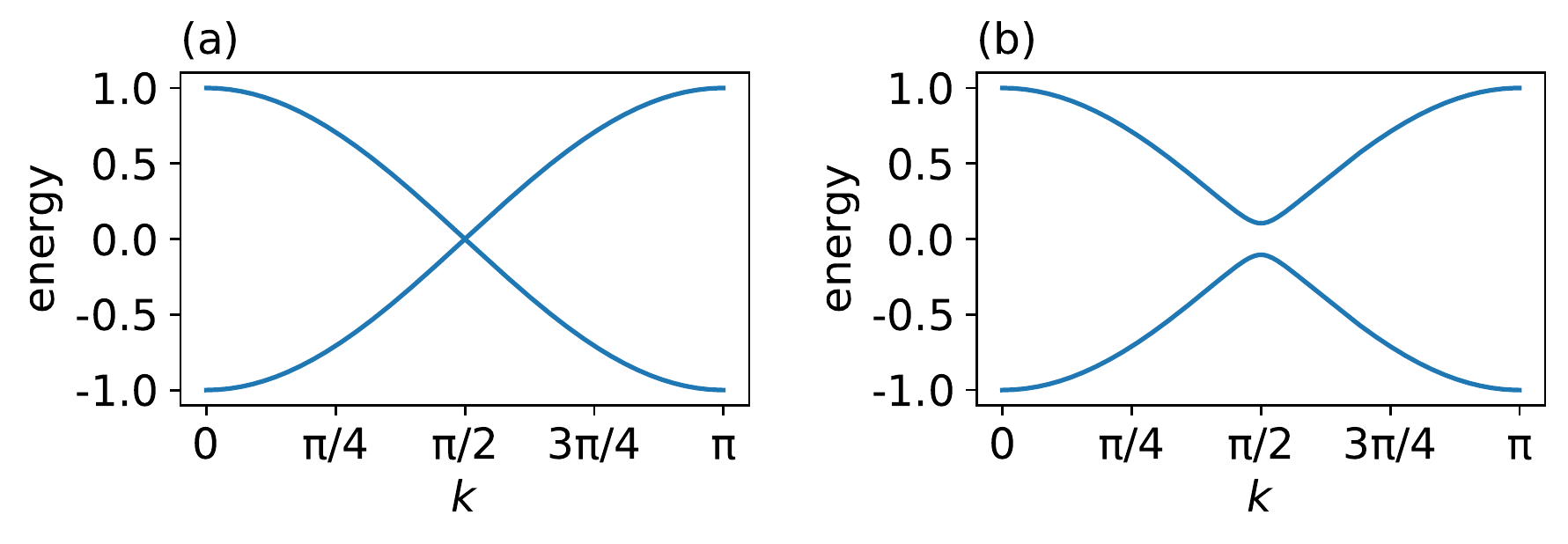}
\caption{
Band structure of Jordan-Wigner fermions
for (a) the inversion-symmetric model $\hxx$
and (b) the inversion-asymmetric model $\hxx+\hstag$
with asymmetric parameters $(\stagXY,\stagZ)=(0.1,0.03)$.
The unit of energy is taken as $J=1$.
}
\label{fig:band}
\efig

%######
We then fermionize the laser-matter couplings and the observables,
and make the $2\times2$-matrix representations in our basis.
The coupling terms are given by
\begin{align}
	\hextZmat(k,t)&= b(t)\sigma_z,\\
	\hextMSmat(k,t) &= e(t) \sin k \sigma_y,
\end{align}
and the electric polarization reads
\begin{align}
	P(k) = -\etasMS\sin k \sigma_y.
\end{align}
The spin current depends on the coupling term,
and its matrix representation is given, for the Zeeman coupling, by
\begin{align}
	\Jspin (k) = J \sin k \sigma_x + \stagXY \cos k \sigma_y,
\end{align}
and, for the magnetostriction effect, by
\begin{align}
	\Jspin (k,t) = J \sin k \sigma_x + [\stagXY-e(t)] \cos k \sigma_y.
\end{align}

We remark that, in the fermion representation,
our model is analogous to two-band models used in the HHG studies for itinerant electrons.
For itinerant electrons, the laser electric field causes the intraband acceleration
and the interband transition, both of which play important roles in the HHG.
In our model, both $\hextZ$ and $\hextMS$ involve the interband and the intraband effects.
To see this, we first consider the special case of $\stagXY=\stagZ=0$,
where $\hfirstmat(k) =  J\cos k \sigma_x$.
Making a unitary transformation $U_0$, we diagonalize this Hamiltonian as $\hfirstmat(k)' =  J\cos k \sigma_z$,
where the coupling terms are represented as 
$\hextZmat(k,t)'= b(t)\sigma_y$ and $\hextMSmat(k,t)' = e(t) \sin k \sigma_x$.
It is manifest that the coupling terms have nonzero elements only in the off-diagonal components,
and thus lead to interband transitions and have no intraband effect.
Next we consider the case of $\stagXY\neq0$ or $\stagZ\neq0$,
where the unitary transformation $U_1$ diagonalizing $\hfirstmat(k)$ is different from $U_0$.
Thus, in the energy eigenbasis, the coupling terms have, in general, diagonal elements,
and some intraband effects are involved.
Although the details such as the $k$-dependence are different,
we expect that the intra- and interband effects in our model result in the HHG.
\hlm{In fact, as we will see at the ends of Secs.~\ref{sec:pol} and \ref{sec:sc},
the harmonic spectra are analogous to those of semiconductors.
}

%##############################
%##############################
%##############################
%#######
\subsection{Time Evolution and Laser Pulse}\label{sec:evolution}
%######
We suppose that the system is initially in the ground state $\gs=\otimes_k\ket{\phi_g(k)}$
and the laser magnetic or electric field is turned off.
In terms of the fermion representation,
the ground state is the one where the lower energy band is fully occupied
and the upper one is completely unoccupied.

%######
The time evolution is caused by either magnetic field $b(t)$ or electric field $e(t)$.
Sufficiently strong field amplitudes ($\sim$ 1MV/cm) at THz regime are obtained for pulse lasers~\cite{Hirori2011,Liu2017,Mukai2016} 
and it is still difficult to generate THz continuous waves with high intensity.
Therefore, we consider $b(t)$ and $e(t)$ of pulse shape:
\begin{align}
	b(t) = b_0 \cos (\Omega t) f(t);
	\quad  e(t) = e_0 \cos (\Omega t) f(t).\label{eq:pulse_field}
\end{align}
Here $b_0$ and $e_0$ are the peak coupling energy,
$\Omega$ is the central frequency,
and $f(t)$ is the Gaussian envelope function,
$f(t)=\exp[-2\ln 2 (t^2/\tfwhm^2)]$,
where $\tfwhm$ represents the full width at half maximum of the intensity $e(t)^2$ or $b(t)^2$.
We refer to $\tfwhm/T$ with $T\equiv 2\pi/\Omega$ as the number of cycles of the pulse field,
which is assumed to be 5 unless otherwise specified below.

%######
As emphasized in Sec.~\ref{sec:intro},
relaxation is important in our setup
because its timescale is comparable to the periods of THZ and GHz waves.
To take account of this effect,
we describe the time evolution by a quantum master equation
of the Lindblad form~\cite{Breuer2007}:
\begin{align}
	\frac{\dd}{\dd t}\rho(k,t)
	&=-\ii [H(k,t),\rho(k,t)] \notag\\
	&\qquad+\gamma \left(L_k\rho(k,t) L_k^\dag -\frac{1}{2}\{ L_k^\dag L_k,\rho(k,t)\}\right),\label{eq:QME}
\end{align}
where $\rho(k,t)$ is the $2\times2$ reduced density matrix
for the subspace with wave number $k$.
Here the Lindblad operator $L_k\equiv\ket{\phi_g(k)}\bra{\phi_e(k)}$
describes the relaxation from the excited state $\ket{\phi_e(k)}$ to the ground state $\ket{\phi_g(k)}$,
and $\gamma$ does its rate.
For simplicity, we assume that $\gamma$ is independent of $k$
and set $\gamma=0.1 J$.
This relaxation rate corresponds to the lifetime $\tau=\gamma^{-1}\sim 7.6\psec$ $(1.5\psec)$
for a typical exchange interaction $J/\kB=10\K$ (50\K).

Our master equation~\eqref{eq:QME} ensures that
the system relaxes to the ground state in the long run
after the external field is switched off.
Without relaxation, the system would remain excited
in an infinitely long time after the pulse irradiation.
We will see later that our master equation approach
thereby allows us to obtain well-defined Fourier spectra
of observables without artificial treatment such as window functions.

In solving the quantum master equation~\eqref{eq:QME},
we take the initial condition $\rho(k,\tini) = \ket{\phi_g(k)}\bra{\phi_g(k)}$
with the initial time $\tini$ $(<0)$ being so small that $f(\tini)\simeq0$.
At time $t$, the expectation value of an observable
\begin{align}
	\hat{O} &= \sum_k \spinorI^\dag \mathcal{O}(k) \spinorI,
\label{eq:Oop}
\end{align}
is given by
\begin{align}
O(t) = \langle \hat{O}\rangle_t= \sum_k \text{tr}[\rho(k,t)\mathcal{O}(k)].	
\end{align}

%##############################
%##############################
%##############################
%##############################
\subsection{Units and Scales of Physical Quantities}\label{sec:units}
Before discussing our results,
we make remarks on the scales of physical quantities.
%######
In the following, we work in the units with $J=1$
and represent all physical quantities including the photon energy, the lifetime of the magnetic excitation,
and the magnetic and the electric fields, in a dimensionless manner with the physical constants set to unity.
The rules to recover the units depend on the value of $J$ that we suppose.
In Table~\ref{tab:units}, we provide the rules for the two choices of $J=10$\,K and 50\,K,
which are typical energy scales of magnets.
In this table, we have assumed that $\etasZ=g\mu_B$ and $\etasMS=g\mu_B/c$,
where $c$ is the speed of light.
The second assumption implies that, in good multiferroic materials,
the magneto-electric coupling is as large as the Zeeman coupling~\cite{Tokura2014,Pimenov2008,Huvonen2009,Furukawa2010}.

%#######
We also provides two tables for convenience.
Table~\ref{tab:unitconv} is the unit conversion table
between different physical quantities.
Table~\ref{tab:flux} shows the correspondence between the electric-(magnetic-)field amplitude
and the energy flux.

%####
\hlm{
As we noted in Introduction,
the maximum intensity of the THz waves
($\sim$1\,MV/cm) is typically smaller than
that of the mid- and near-infrared lasers used in HHG measurements in semiconductors.
%Also, the excitation energy in magnets is smaller than that in semiconductors.
Table~\ref{tab:units} tells us that
this corresponds to $\sim0.1J$ at most.
Therefore we will mainly focus on relatively lower-order
($n=2,3,4,5$) harmonics in Secs.~\ref{sec:pol}--\ref{sec:experiments}.
}

%##############################
%##############################
%##############################
%##############################
\section{Electric Polarization}\label{sec:pol}
In this section, we discuss the high-harmonic spectrum
of the electric polarization.
We use the Hamiltonian $\hfirst$~\eqref{eq:Hfirst},
and discuss the effects of driving by either $\hextZ$~\eqref{eq:HextZ} or $\hextMS$~\eqref{eq:HextMS}.
As discussed in Sec.~\ref{sec:units}, we work in the dimensionless units
corresponding to Table~\ref{tab:units}.

\begin{table}[h]
  %###########
  	\caption{Table of units for physical parameters depending on two choices of $J=10$\,K and 50\,K.}
  \begin{tabular}{l r r}
   Energy, $J$ & 10\,K & 50\,K \\
   \hline\hline
   Photon energy, $\hbar\Omega$ & 0.86\,meV & 4.3\,meV\\
   Time, $\hbar/J$ & 0.76\,ps & 0.15\,ps\\
   Frequency, $f=\Omega/(2\pi)$ &  0.21\,THz   &  1.0\,THz \\
   Magnetic field, $B_0=J/(g\mu_B)$ & 7.4\,T & 37\,T \\
   Electric field, $E_0=cB_0$ & 22\,MV/cm & 112\,MV/cm\\
   \hline\hline
  \end{tabular}
    \label{tab:units}
  %############
  	\caption{Unit conversion table.}
  	\begin{tabular}{l r r}
   EM field & THz & GHz\\
   \hline\hline
   Frequency, $\Omega/(2\pi)$ & $10^{12}$\,Hz & $10^{9}$\,Hz\\
   Energy, $\hbar\Omega$ & 4.1\,meV& 4.1\,$\mu$eV \\
   Temperature, $T=\hbar\Omega/k_\text{B}$ & 48\,K & 48\,mK \\
   Magnetic field, $B_0=\hbar\Omega/g\mu_\text{B}$ & 36\,T & 36\,mT\\
   Electric field, $E_0=cB_0$ & 107\,MV/cm &  107\,kV/cm\\
   \hline
   \hline
  \end{tabular}
  \label{tab:unitconv}
  %##############
  \caption{Laser energy flux for reference field strengths.}
	\begin{tabular}{cc}
	 & $E_0=$1\,MV/cm  \\
	   \hline\hline
	   Magnetic field, $B_0$ & 0.33\,T \\
	   Energy flux, $I$ & 1.3\,GW/cm${}^2$\\
	   \hline\hline
	\end{tabular}
	\label{tab:flux}
\end{table}

%###################
\bfig
\putfigraw{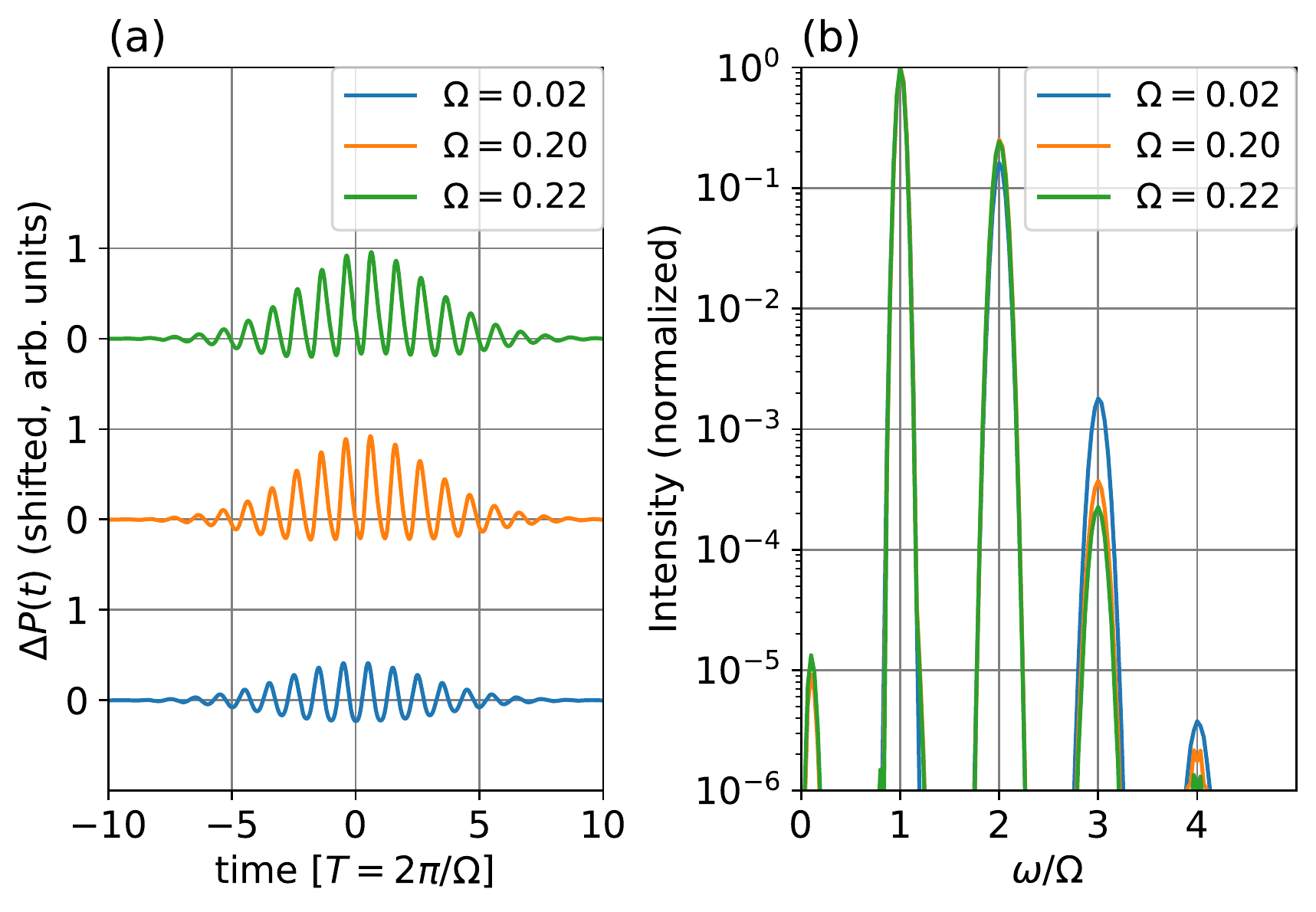}
\caption{
(a) Time profile of $\dP(t)$ for the Zeeman driving $\hextZ(t)$
with frequency $\Omega=0.02$, 0.2, and 0.22.
The parameters are $\stagXY=0.1$, $\stagZ=0.03$, and $b_0=0.02$,
where the spin gap is $\gapI=0.21$.
(b) Corresponding power spectrum $I_P(\omega)$~\eqref{eq:powP}
plotted against $\omega/\Omega$.
}
\label{fig:spec_pol_Z}
\efig

%###### omega > gap , omega <gap
%###### real time ######
We first investigate the typical behaviors 
of the time profile $\dP(t)$ and the corresponding power spectrum $I_P(\omega)$~\eqref{eq:powP}
obtained by $\hextZ$~\eqref{eq:HextZ}.
Figure~\ref{fig:spec_pol_Z} shows the results obtained for the parameters
$\stagXY=0.1$, $\stagZ=0.03$, and $B_0\etasZ=0.02$, with several driving frequencies.
Here the power spectrum is normalized so that the fundamental harmonic $I_P(\Omega)$ is unity.
We note that the even-order harmonics are present since the inversion symmetry is broken now.

%###
The lowest frequency $\Omega=0.02$,
which is approximately 10 times smaller than the spin gap $\gapI=0.21$,
corresponds to the standard setup for the semiconductor HHG
and the previous study of spin-system HHG~\cite{Takayoshi2019}.
At this lowest frequency, strong harmonic peaks are obtained
and the peak heights slowly decrease as the harmonic order increases.
At this frequency, $\Omega\ll \gamma^{-1}$ holds,
and thus the dynamics is nearly adiabatic.
Namely, relaxation occurs so fast that the quantum state
always approaches to ground state at the instantaneous external field.

%######
The harmonic peaks are present regardless of
whether the driving frequency is well below the gap
or near-resonant.
In fact, in Fig.~\ref{fig:spec_pol_Z}(b),
we find strong harmonic peaks for $\Omega=0.20$ and $0.22$,
which are slightly below and above the spin gap $\gapI=0.21$, respectively.
At these frequencies $\Omega\sim \gamma^{-1}$,
the quantum dynamics is more coherent than that for $\Omega=0.02$ 
(i.e., less suffers from the environment), but the relaxation is still effective to
keep harmonic peaks strong and sharp.
Regarding experiments with THz laser pulse,
it is advantageous that the harmonics peaks are seen with higher frequencies
because the spin gap is typically smaller than 1THz-photon energy in many of magnetic insulators 
and it becomes more difficult to obtain high field amplitudes
in the frequency regime lower than 1THz.

%###################
\bfig
\putfigraw{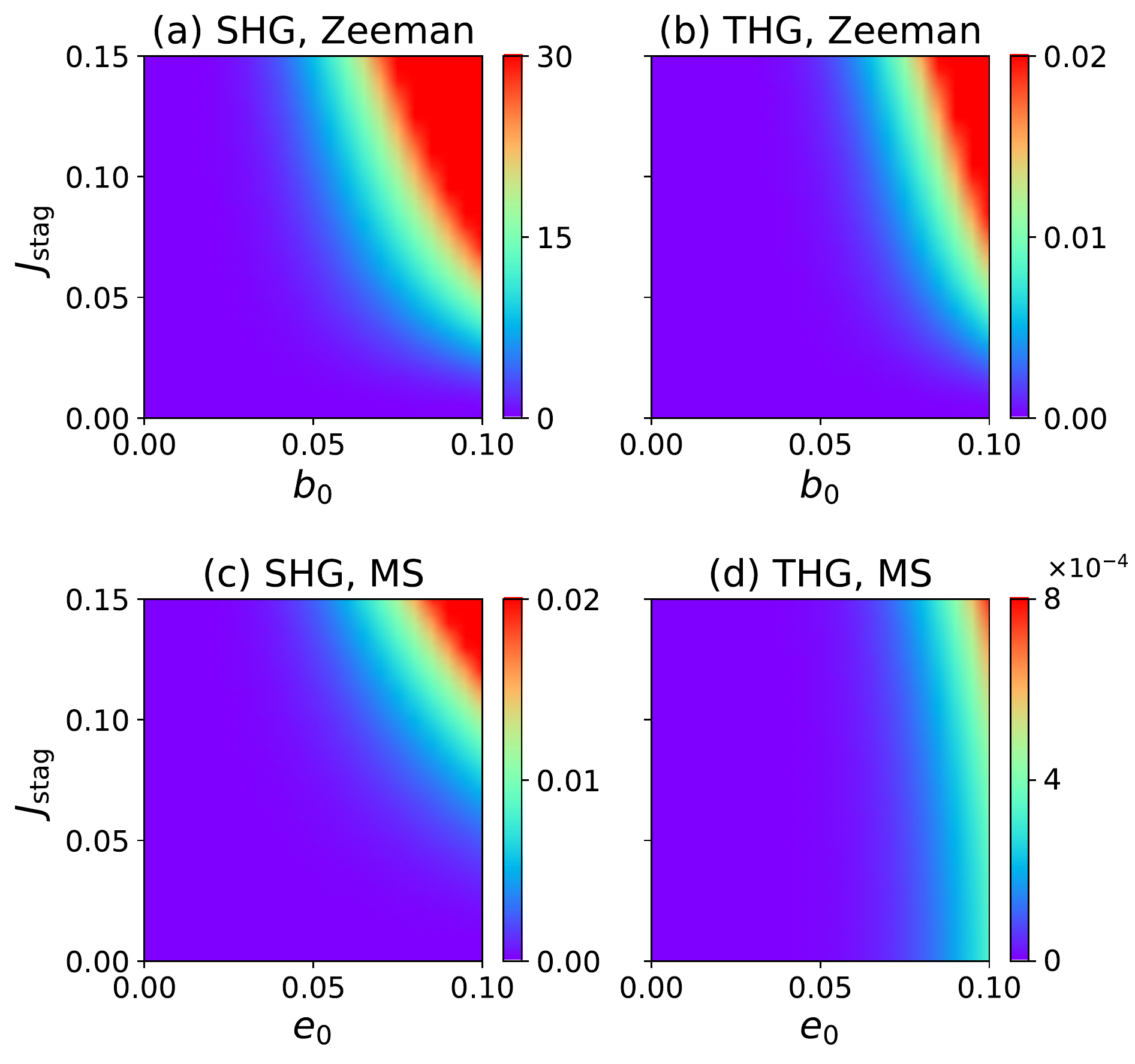}
\caption{
Intensities of SHG [$I_P(2\Omega)$] (left) and THG [$I_P(3\Omega)$] (right)
from electric polarization
driven by ac Zeeman coupling (upper) and magnetostriction effect (lower)
at $\Omega=0.5$.
The unit of intensity in panels (a) and (b)
is chosen as $I_P(\Omega)$ for $b_0=\stagXY=0.05$,
and that in panels (c) and (d)
as $I_P(\Omega)$ for $e_0=\stagXY=0.05$.
The other parameters are set as $\gamma=0.1$ and $\stagZ=0.03$.
The ratio of the two units,
$I_P(\Omega)_{b_0=\stagXY=0.05}/I_P(\Omega)_{e_0=\stagXY=0.05}$, is $1.4\times10^{-3}$.
Namely, the fundamental harmonic for the magnetostriction
is much larger than that for the Zeeman coupling.
}
\label{fig:color_pol}
\efig

%###########
Now we systematically investigate the intensity of the second- and third-harmonic generation (SHG and THG).
We fix $\stagZ=0.03$ and $\Omega=0.5$,
and calculate the harmonic spectrum for various sets of parameters $(b_0,\stagXY)$.
Figures~\ref{fig:color_pol}(a) and (b) respectively show $I_P(2\Omega)$ and $I_P(3\Omega)$
driven by the ac Zeeman coupling $\hextZ(t)$.
Here the unit of the intensity is chosen as the fundamental harmonic $I_P(\Omega)$ for $b_0=\stagXY=0.05$.
Both the SHG and the THG tend to increase as $b_0$ or $\stagXY$ increases,
and these signals can be as large as the fundamental harmonic. 
It is very natural that the HHG signal grows up with the increase of the light-spin coupling $b_0$. 
Moreover, the growth of SHG with the increase of $J_{\rm stag}$ is easily understood 
because the inversion symmetry is broken by the presence of both $J_{\rm stag}$ and $H_{\rm stag}$, 
and the SHG disappears in inversion-symmetric systems. 
In fact, we have confirmed that in the limit of $\stagXY\to 0$, where the site-center inversion symmetry recovers,
the SHG vanishes in line with the selection rule.

HHG is also obtained by the magnetostriction effect $\hextMS(t)$.
Figures~\ref{fig:color_pol}(c) and (d) respectively show $I_P(2\Omega)$ and $I_P(3\Omega)$ in the $(e_0,\stagXY)$ plane,
where we again fix $\stagZ=0.03$ and $\Omega=0.5$.
As we already mentioned,
the coupling constant $e_0$ can be as larger as the ac Zeeman one $b_0$
in multiferroic magnets, and we thereby set the maximum value of $e_0$
to be that of $b_0$ in Fig.~\ref{fig:color_pol}.
Compared with (a) and (b),
the panels (c) and (d) seem to imply that the SHG and the THG
by the magnetostriction effect are much smaller
than those by the ac Zeeman coupling.
However, this is mainly because
the fundamental harmonic $I_P(\Omega)$ is quite large
for the magnetostriction case,
and the absolute values of the SHG and the THG
are comparable in the two cases. 

%#####
In addition to $\Omega=0.5$, we have investigated
$\Omega=0.2$, which is closer to the spin gap (data not shown).
In this case, the signals tend to become large when
the spin gap approaches the photon energy $\hbar\Omega$.
In this parameter region,
we have confirmed that the absolute values
of the SHG and the THG are somewhat larger
for the magnetostriction effect
than for the ac Zeeman coupling.

%######
Let us estimate the required laser-field amplitudes to observe HHG in experiments
(see Sec.~\ref{sec:experiments} for experimental protocols).
Considering a recent experiment in an antiferromagnetic crystal~\cite{Lu2017},
we suppose that a 1\% intensity (10\% amplitude)
of the fundamental harmonic is actually detectable.
We apply this criterion to our calculations
for $\stagXY=0.05$ and $\stagZ=0.03$ at $\Omega=0.5$,
which corresponds to $f=0.10$\,THz (0.52\,THz)
for magnets of energy scale $J=10$\,K (50\,K) (see Table~\ref{tab:units}).
As for the ac Zeeman driving with, e.g., $b_0=0.01$,
the SHG has about a 10\% intensity of the fundamental harmonic
and should be observable
whereas the THG does about a $10^{-4}$ intensity
and its detection might be challenging.
Thus we regard $b_0=0.01$ as a required field amplitude
to observe HHG in experiments.
According to Table~\ref{tab:units},
this field amplitude corresponds to $B_0=74$\,mT (370\,mT)
for magnets of energy scale $J=10$\,K (50\,K).
From Table~\ref{tab:flux}, this magnetic field amplitude corresponds to
the electric field amplitude $E_0=220$\,kV/cm (1.1\,MV/cm)
and the energy flux $I=0.29$\,GW/cm${}^2$ (1.5\,GW/cm${}^2$).
These field amplitudes are within the reach of 
the current THz-laser technology~\cite{Hirori2011,Dhillon2017,Liu2017,Mukai2016}.
As for the magnetostriction effect, the required amplitude is larger than the above values by some factor.

%##############################
%##############################
%##############################
%##############################
\section{Spin Current}\label{sec:sc}
%###################
In this section, we investigate the harmonic spectrum of the spin current.
As in the previous section, we use the Hamiltonian $\hfirst$~\eqref{eq:Hfirst}
and consider the effects of driving by either $\hextZ$~\eqref{eq:HextZ} or $\hextMS$~\eqref{eq:HextMS}.

\bfig
\putfigraw{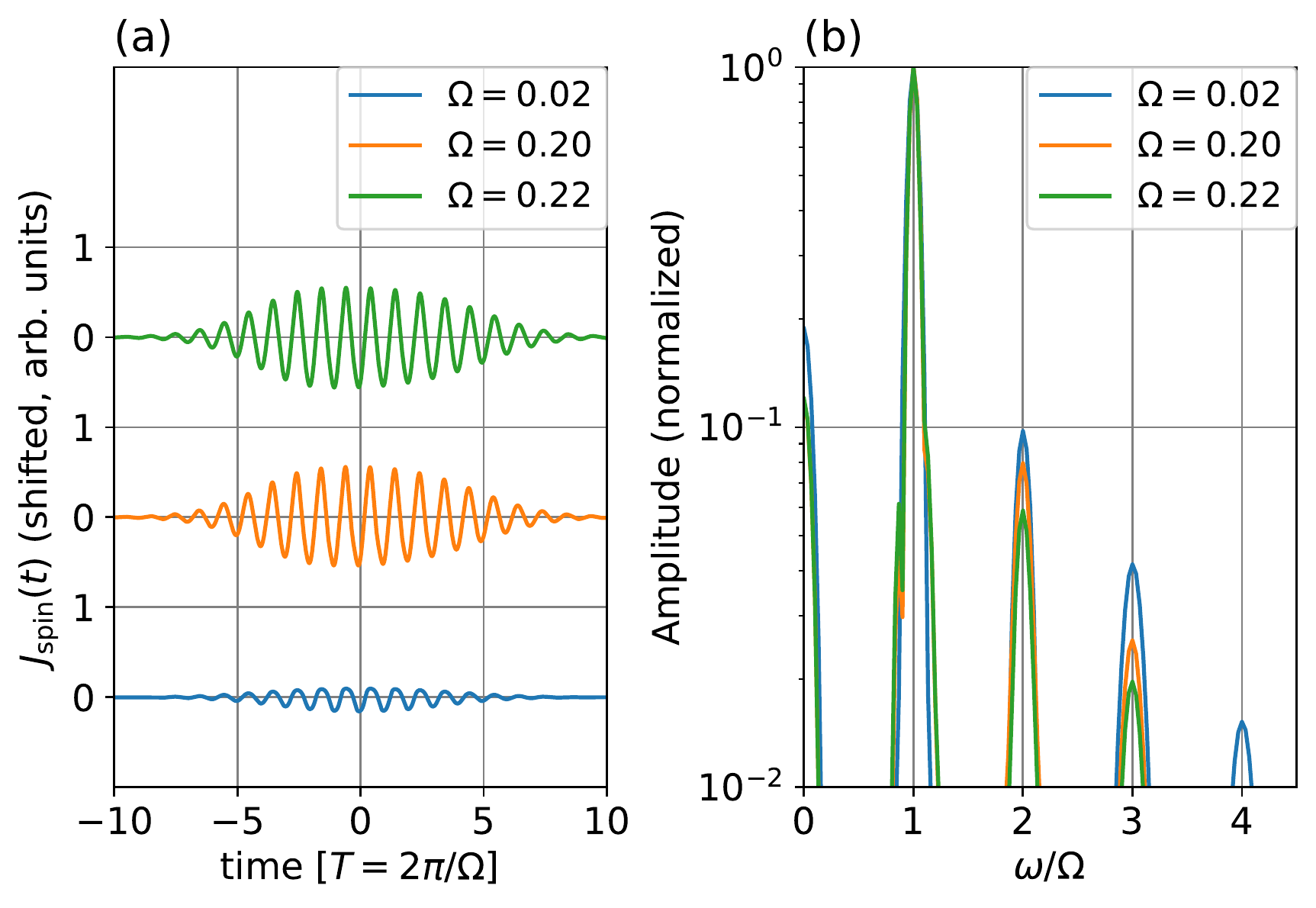}
\caption{
(a) Time profile of $\Jspin(t)$ for the ac Zeeman driving $\hextZ(t)$
with frequency $\Omega=0.02$, 0.2, and 0.22.
The parameters are $\stagXY=0.1$, $\stagZ=0.03$, and $b_0=0.1$,
where the spin gap is $\gapI=0.21$.
(b) Corresponding \hlr{amplitude} spectrum $|\Jspin(\omega)|$~\eqref{eq:powP}
plotted against $\omega/\Omega$.
}
\label{fig:spec_sc_Z}
\efig

%######
The ac Zeeman driving $\hextZ$~\eqref{eq:HextZ}
gives rise to harmonic peaks in the spin-current spectrum $|\Jspin(\omega)|$
similarly to the electric polarization.
Figure~\ref{fig:spec_sc_Z} shows the typical time profile $\Jspin(t)$
and spectrum $|\Jspin(\omega)|$,
where the parameters are the same as in Fig.~\ref{fig:spec_pol_Z}.
Here the spectrum is normalized so that the fundamental harmonic $|\Jspin(\Omega)|$ is unity.
Again, the clear peaks are observable both for $\omega=0.02\sim 0.1\gapI$ and $\omega \sim \gapI$.

%######
We note that there also exists the dc ($\omega=0$) component of the spin current.
As was proposed in Refs.~\cite{Ishizuka2019a,Ishizuka2019b}, this corresponds to the rectification of the spin current,
which occurs in inversion-asymmetric magnets.
For $\Omega=0.2$ and $0.22$, the peak heights of $\omega=0$ and $\omega=2\Omega$
are similar as shown in Fig.~\ref{fig:spec_sc_Z}(b).
Thus, if the spin-current rectification is observed in an experiment,
the second harmonic is also likely to be observed in the experimental setup.

%#####
\bfig
\putfigraw{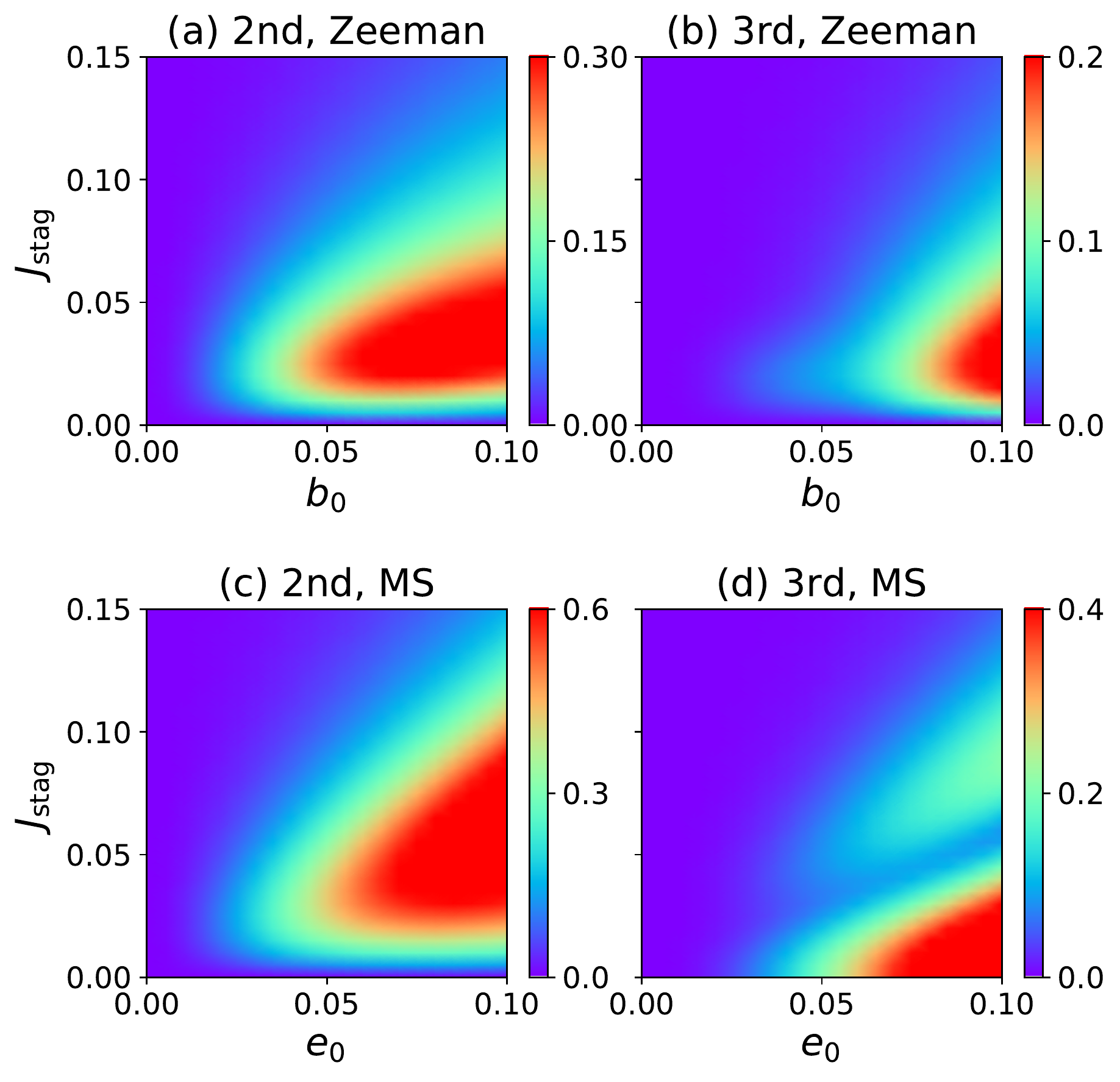}
\caption{
Intensities of second- [$|\Jspin(2\Omega)|$] (left) and third-harmonic spin currents [$|\Jspin(3\Omega)|$] (right)
by the ac Zeeman coupling (upper) and the magnetostriction effect (lower).
The unit of intensity is chosen as $|\Jspin(\Omega)|$ for $b_0$ $(e_0)$ $=\stagXY=0.05$.
The other parameters are set as $\gamma=0.1$, $\stagZ=0.03$, and $\Omega=0.02$.
}
\label{fig:color_sc}
\efig

%#######
The peak heights of the second and the third harmonics
are systematically shown in Fig.~\ref{fig:color_sc}
for the ac Zeeman coupling and the magnetostriction effect
as was done for the electric polarization in the previous section.
Here we use a lower frequency $\Omega=0.02$, which corresponds to $f=4.2$\,GHz (21\,GHz)
for magnets of energy scale $J=10$\,K (50\,K) (see Table~\ref{tab:units}).
This lower frequency is advantageous in experimentally observing the spin currents
in electric circuits because the upper limit of the frequency $\Omega$ for the electric detection 
is in a GHz regime~\cite{Wei2014}. In Sec.~\ref{sec:experiments}, we will discuss in detail 
how high harmonic spin currents are observed by electric techniques. 

The second harmonic spin current
exhibits a nonmonotonic behavior as a function of $\stagXY$
while monotonically increases with $b_0$ or $e_0$
as shown in the panels (a) and (c).
This nonmonotonic behavior arises from
the following two limits. First, the second harmonic vanishes
in the limit $\stagXY\to0$ owing to the inversion symmetry.
Second, it tends to decrease for the larger $\stagXY$.
This can be understood by the analogy between HHG of semiconductors and that of the present spin systems.
Namely, the photon energy $\hbar\Omega$ at $\Omega=0.02$ is much smaller than the spin gap $\Delta_{\rm I}$ 
for a sufficiently large $J_{\rm stag}$ and thus nonlinear optical processes such as multi-photon absorption 
is necessary for generating magnetic excitations and spin dynamics. 
Such nonlinear dynamics becomes suppressed in systems with a large $J_{\rm stag}$ (i.e., a large spin gap) 
and therefore the height of the second harmonic spin current decreases in the large-$J_{\rm stag}$ region.  
We remark that the third harmonics also exhibits the similar behavior to the second one, 
but more nontrivial $J_{\rm stag}$ dependence can arise in the third harmonics  
as shown in the panels (b) and (d) of Fig.~\ref{fig:color_sc}. 
In particular, there exist a clear double peak structure in the panel (d). 
The authors do not have simple interpretation for this complicated behavior yet.

In addition to the case of $\Omega\ll \Delta_{\rm I}$, 
we have also studied the second- and third-harmonic spin currents at $\Omega\sim\Delta_{\rm I}$. 
In this case, the intensity profiles of $|\Jspin(2\Omega)|$ and $|\Jspin(3\Omega)|$ 
in the $(b_0(e_0),J_{\rm stag})$ plane are respectively similar to those of 
$|I_{P}(2\Omega)|$ and $|I_{P}(3\Omega)|$ in Fig.~\ref{fig:color_pol}.

%######
Now we discuss the typical field amplitude required in experimental observations,
following the same criterion mentioned at the end of Sec.~\ref{sec:pol}.
As shown in Fig.~\ref{fig:spec_sc_Z},
the second harmonic spin current has about a 10\%
of the fundamental one at $b_0=0.1$,
and thus we regard this amplitude as required.
According to Table~\ref{tab:units},
this field amplitude corresponds to $B_0=0.74$\,T (3.7\,T)
for magnets of energy scale $J=10$\,K (50\,K).
From Table~\ref{tab:flux}, this magnetic field amplitude corresponds to
the electric field amplitude $E_0=2.2$\,MV/cm (11\,MV/cm)
and the energy flux $I=0.29$\,GW/cm${}^2$ (1.5\,GW/cm${}^2$).
The required amplitude for the magnetostriction effect
is the same as the above values.
The concrete experimental setups will be discussed in
Sec.~\ref{sec:experiments}.

%##############################
%##############################
%##############################
%##############################
\section{Magnetization}\label{sec:mag}
We have discussed the harmonic generation and spin current so far
by using Hamiltonian~\eqref{eq:Hfirst},
in which the total magnetization is conserved.
For completeness, in this section, we switch to another Hamiltonian (see Eq.~\eqref{eq:Hsecond} below),
investigating the harmonic generation through nonlinear magnetization dynamics.
The methods that we have developed in previous sections
apply to this model.

\subsection{Model and Formulation}\label{sec:mag_model}
%#########
The second Hamiltonian that we consider in this section
is the anisotropic XY model:
\begin{align}\label{eq:Hsecond}
    \hsecond=\hxy &= J \sum_{j=1}^L \left[ (1+\aniso)\hS_j^x \hS_{j+1}^x 
\nonumber\right.\\
&\left. \qquad\qquad+ (1-\aniso)\hS_j^y \hS_{j+1}^y
-\beta_u\hat{S}^z_j \right],
\end{align}
where $\aniso$ quantifies intraplane anisotropy and the last term with $\beta_u=g\mu_B B_0$ 
represents the uniform Zeeman coupling due to an applied external magnetic flux $B_0$.
For $\aniso\neq0$, the total magnetization $\sum_{j}\hS_j^z$ is not a conserved quantity,
and the Hamiltonian~\eqref{eq:Hsecond} is useful to study dynamics of magnetization.
The case of the strongest Ising anisotropy, i.e., $\epsilon=1$ corresponds to 
the so-called transverse-field Ising model. 
There exist several quasi-one-dimensional magnets with strong Ising anisotropy, 
e.g., CoNb${}_2$O${}_6$~\cite{Coldea2010}, BaCo${}_2$V${}_2$O${}_8$~\cite{Kimura2007a,Kimura2008}, 
and SrCo${}_2$V${}_2$O${}_8$~\cite{Wang2018}.

We consider the laser-spin interaction by 
the ac Zeeman coupling to the laser magnetic field $B(t)$ along the $S^z$ direction
as we have done in Sec.~\ref{sec:model1}.
Note that we do not consider staggered effects in this section.
Thus the coupling Hamiltonian is given by
\begin{align}\label{eq:HextZ2}
\hextZnew(t) = -\beta(t) \sum_{j}\hS_j^z
\end{align}
with $\beta(t)\equiv B(t)\etauZ$.

%####
The total magnetization
\begin{align}\label{eq:magnetization}
	\hM = \sum_j \hS^z_j.
\end{align}
is the observable that we consider for $\hsecond+\hextZnew(t)$.
As is the case with the electric polarization,
the magnetization becomes the source of electromagnetic radiation when varies in time.
Thus we consider the radiation power %at frequency
\begin{align}\label{eq:powM}
	I_M(\omega) \propto |\omega^2 M(\omega)|^2,
\end{align}
and discuss its peak structure in the following.
As we remarked for the electric polarization in Sec.~\ref{sec:model1},
a constant shift of $M(t)=\langle \hat{M}\rangle_t$ does not change $I_M(\omega)$,
and thus we may use $\varDelta M(t)=M(t)-M_0$
with $M_0=\langle \hat M\rangle_{\tini}$.

\hlm{We remark that the odd-order harmonics
exist for generic choices of the parameters
whereas the even-order ones appear only when $\beta_u\neq0$.
This is analogous to the HHG selection rule
in semiconductors regarding the inversion
that has been discussed in Sec.~\ref{sec:setup1}.
Note, however, that the selection rule for $I_M(\omega)$
does not follow from the inversion symmetry
\hlm{unlike the HHG in semiconductors~\cite{Alon1998}}
since the magnetization is even under the inversion.
For the magnetization,
the rule can be obtained by using spin rotations as follows.
}
\hlr{
We consider the ideal situation where
$\beta(t)$ involves many cycles
and is approximately sinusoidal
with period $T=2\pi/\Omega$.
}
For the special case of $\beta_u=0$,
the total Hamiltonian $\hsecond+\hextZnew(t)$
is invariant under a dynamical transformation given by
$t\to t+T/2$ combined with the global $\pi$ rotation around, e.g., the $S^y$ axis.
Since $\hat{M}$ is odd under this transformation,
\hlr{we have $M(t+T/2)=-M(t)$,
which implies that the even-order HHG is prohibited.
In fact, we have $M(2n\Omega)\propto \int_0^{T}\dd t\, \ee^{\ii 2n\Omega t}M(t)=\int_0^{T}\dd t\,\ee^{\ii 2n\Omega t}[M(t)+M(t+T/2)]/2=0$
(see Appendix~\ref{app:ds} for more detail).}
However, for $\beta_u\neq0$, this dynamical symmetry is broken
and the even-order HHG by magnetization is allowed
even if the inversion symmetry is present.
%\hlr{We remark that the above argument assumes $\beta(t)$ of multi cycles and breaks down for very short, e.g., sub-cycle external fields.}

%To put the above argument in a different way,
\hlm{The above result shows that}
the even-order HHG can be controlled by
the static magnetic field $\beta_u$,
and this is a close analogy to the SHG controlled by the electric current in semiconductors~\cite{Ruzicka2012}
and superconductors~\cite{Moor2017,Nakamura2019}.
We stress that this controllability applies to a wide class
of spin systems as long as the dynamical symmetry exists
in the absence of the static magnetic field.

%######
\subsection{Fermionization and BCS Hamiltonian}
Our new model is also fermionized via the Jordan-Wigner transform. 
Through the Fourier transformation of the Jordan-Wigner fermion 
$\hat{d}_k \equiv L^{-1/2}\sum_{j=1}^L \ee^{-\ii k j} \ann_{j}$,
the Hamiltonian~\eqref{eq:Hsecond} and the coupling~\eqref{eq:HextZ2} are given by 

\begin{align}
	&\hsecond = \sum_{k} \left [
		(J\cos k+\beta_u) \hat{d}^\dag_k \hat{d}_k\right.\notag\\
		&\qquad\qquad\qquad\left.+\ii (J/2)\epsilon \sin k (\hat{d}^\dag_{k}\hat{d}^\dag_{-k}
		+\hat{d}_{k}\hat{d}_{-k})\right],\label{eq:HII_fer}\\
	&\hextZnew(t) = \beta(t)\sum_k \hat{d}_k^\dag \hat{d}_k.\label{eq:HextZ2_fer}
\end{align}
For simplicity, we assume that $L$ is even
and focus on the subspace of the states with even fermion numbers.
Correspondingly, we impose the anti-periodic boundary condition
for the fermion and
take $k=\pm\pi (m+1/2)/L$ $(m=0,1,\dots,L-1)$~\cite{Chakrabarti1996}.

Equation~\eqref{eq:HII_fer} is the same as a BCS-type Hamiltonian for superconductors~\cite{Schrieffer1971,Tinkham2004}. 
It means that the HHG of the anisotropic spin chain~(\ref{eq:Hsecond}) is analogous to 
that of superconducting systems with a static Cooper-pairing coupling $\ii\epsilon J\sin k/2$ 
(i.e., without the dynamics of condensed wave function such as the Higgs mode). 
Similarly to Eq.~(\ref{eq:H1}), 
we have the $2\times 2$ form of the Hamiltonian~(\ref{eq:HII_fer}) as
\begin{align}
	\hsecond = \sum_{k>0}\phi^\dag_{{\rm II},k} \hsecondmat(k)\phi_{{\rm II},k}\label{eq:H2_2by2}
\end{align}
where $\phi_{{\rm II},k}\equiv {}^{\text{t}}(\hat d_k,\hat d_{-k}^\dag)$ and the $2\times2$ Hamiltonian 
matrix $\hsecondmat(k)$ is defined by 
\begin{align}
	\hsecondmat(k) = (J\cos k+\beta_u) \sigma_z -J\epsilon \sin k \sigma_y.
\label{eq:hIImat}
\end{align}
To diagonalize Eq.~\eqref{eq:H2_2by2},
we perform a unitary (Bogoliubov) transformation.
Namely, we introduce $\gamma_k = u_k \hat d_k -v_k \hat d_{-k}^\dag$ and $\gamma_{-k} = v_k \hat d_k^\dag +u_k \hat d_{-k}$,
where $u_k=\cos(\theta_k/2)$ and $v_k=\ii\sin(\theta_k/2)$ with $\tan\theta_k=-J\aniso\sin k/(J\cos k+\beta_u)$, obtaining
$\hsecond = \sum_{k>0}\epII(\hat \gamma_k^\dag \hat \gamma_k + \hat \gamma_{-k}^\dag \hat \gamma_{-k}-1)$
%\begin{equation}
%	\hsecond = \sum_{k>0}\epII(\hat \gamma_k^\dag \hat \gamma_k + \hat \gamma_{-k}^\dag \hat \gamma_{-k}-1)
%	\label{eq:H2_Bogo}
%\end{equation}
with $\epII=J\sqrt{(\cos k+\tfig)^2+(\epsilon\sin k)^2}$
and $\tfig=\beta_u/J$.
Thus, the ground state is the one annihilated
by all the $\hat \gamma_{\pm k}$'s
and written as~\footnote{In the case of $|\tfig|\le1$,
in addition to $\gsII$ of Eq.~\eqref{eq:gsII},
there exists an almost degenerate state in the subspace
of states with odd fermion numbers:
$\gsIIp = \hat{d}_\pi^\dag \prod_{0<k<\pi} (u_k +v_k \hat{d}_k^\dag \hat{d}_{-k}^\dag) \ket{0}$
with $k=\pi m/L$ $(m=1,2,\dots,L-1)$.
These states $\gsII$ and $\gsIIp$ are eigenstates of the global $\pi$ rotation around the $S^z$ axis $\hat{U}^z_\pi$, $\hat{U}^z_\pi\hS_j^\alpha\hat{U}^{z\dag}_\pi=-\hS_j^\alpha$ $(\alpha=x,y)$, with eigenvalue 1 and $-1$, respectively.
Meanwhile, the two N\'{e}el
states $\ket{\mathcal{N}_+}=\ket{\rightarrow\leftarrow\rightarrow\cdots}$ and $\ket{\mathcal{N}_-}=\ket{\leftarrow\rightarrow\leftarrow\cdots}$ in the thermodynamic limit satisfy $\hat{U}^z_\pi\ket{\mathcal{N}_\pm}=\ket{\mathcal{N}_\mp}$.
Thus $\ket{\Psi_\pm}\equiv (\gsII\pm\gsIIp)/\sqrt{2}$ satisfying $\hat{U}^z_\pi\ket{\Psi_\pm}=\ket{\Psi_\mp}$ correspond to the N\'{e}el states.
Nevertheless, for large system sizes, all these states lead to almost the same
dynamics for $\hat{M}_z$ because $\braket{\Psi_\text{II}'|\hat{M}_z(t)|\Psi_\text{II}}=0$
and $\braket{\Psi_\text{II}|\hat{M}_z(t)|\Psi_\text{II}}\simeq\braket{\Psi_\text{II}'|\hat{M}_z(t)|\Psi_\text{II}'}$,
where $\hat{M}_z(t)$ is the Heisenberg-picture operator
}
\begin{align}
	\gsII =   \prod_{k>0} (u_k +v_k \hat{d}_k^\dag \hat{d}_{-k}^\dag) \ket{0},
\label{eq:gsII}
\end{align}
where $\ket{0}$ being the Fock vacuum for the fermion $\{\hat d_k\}$.

Even under the ac Zeeman coupling~\eqref{eq:HextZ2_fer},
the time evolution of each $k$-subspace
occurs within a two-dimensional space
rather than the entire four-dimensional space.
This is because the coupling conserves
the number of the Jordan-Wigner ($d$) fermions
and a single quasiparticle ($\gamma$) excitation
is prohibited.
We let $\hat d^\dag_k \hat d^\dag_{-k}\ket{0_k}$ and $\ket{0_k}$ ($\ket{0_k}$ is the Fock vacuum for the $k$-subspace) 
be the basis then the $2\times2$-matrix representation
of $\hsecond$ is given by Eq.~\eqref{eq:hIImat} whose
eigenstates are $\ket{\psi_g(k)}={}^{\text{t}}(v_k,u_k)$ and $\ket{\psi_e(k)}={}^{\text{t}}(u_k,v_k)$.
The two eigenenergies $\pm\epII$ define the two energy bands.

%###################
\bfig
\putfigraw{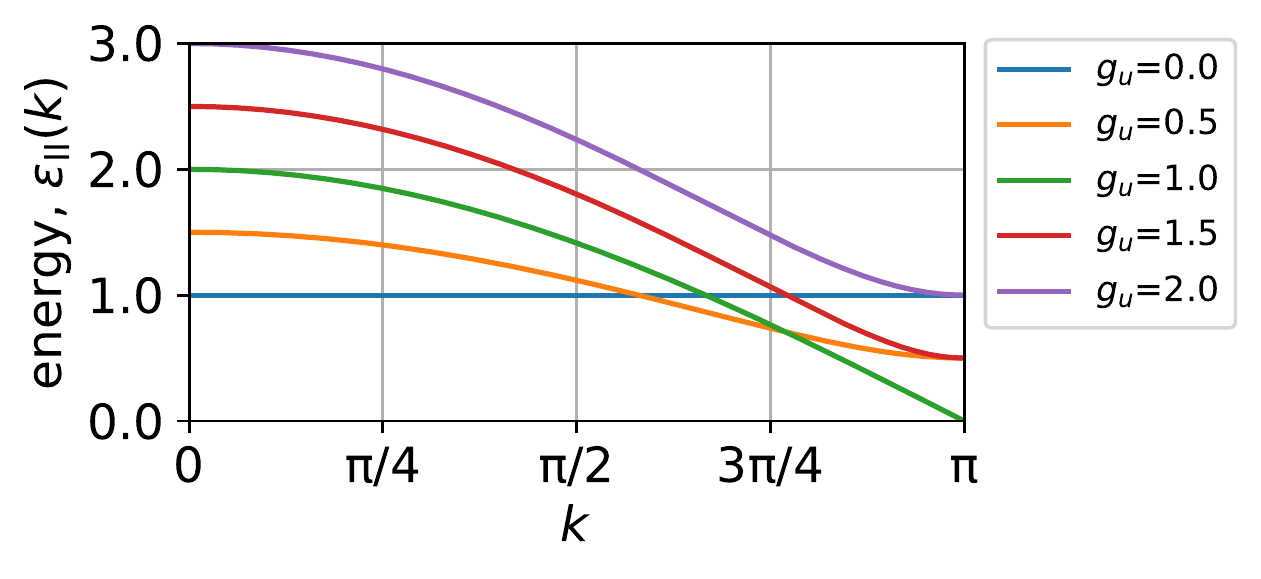}
\caption{
Upper band $\epII$ of the $2\times2$ matrix~\eqref{eq:HIImat} for the transverse-field Ising chain ($\epsilon=1$)
in static Zeeman fields $\tfig=0.0,0.5,\dots$, and 2.0.
The energy unit is $J$ and the lower band is given by $-\epII$.
}
\label{fig:band_TFI}
\efig

In the following, we focus on $\epsilon=1$
corresponding to the transverse-field Ising model
and assume $\beta_u>0$.
The energy bands are illustrated in Fig.~\ref{fig:band_TFI}.
Then the energy gap, i.e., the minimum energy difference
between the upper and lower bands,
occurs at $k=\pi$ and is given by
\begin{align}
	\gapII = 2J|1-\tfig|.\label{eq:gapII}
\end{align}
In terms of the original spin model,
$\tfig=1$ corresponds to the quantum critical point
between the N\'{e}el ($0<g_{\rm u}<1$) and the forced ferromagnetic ($g_{\rm u}>1$) 
phases~\cite{Chakrabarti1996,Sachdev2011}.

Since our problem is two-dimensional in the above sense,
we can use the quantum master equation~\eqref{eq:QME}
to analyze the dynamics with relaxation.
In the present case,
the Hamiltonian part corresponds to
\begin{align}
	H(k,t) = \hsecondmat(k)+\beta(t)\sigma_z.\label{eq:HIImat}
\end{align}
and the Lindblad operator
is $L_k=\ket{\psi_g(k)}\bra{\psi_e(k)}$.
The master equation is solved for the density matrix
with the initial condition
$\rho(k,t_\text{ini})=\ket{\psi_g(k)}\bra{\psi_g(k)}$.
We set the driving field as a pulse shape
\begin{align}
\label{eq:pulse_mag}
	\beta(t) = \beta_0 \cos (\Omega t) f(t), 
\end{align}
where $\beta_0$ is the peak coupling energy
and $f(t)$ is the same as that in Sec.~\ref{sec:model1}.

%###################
\bfig
\putfigraw{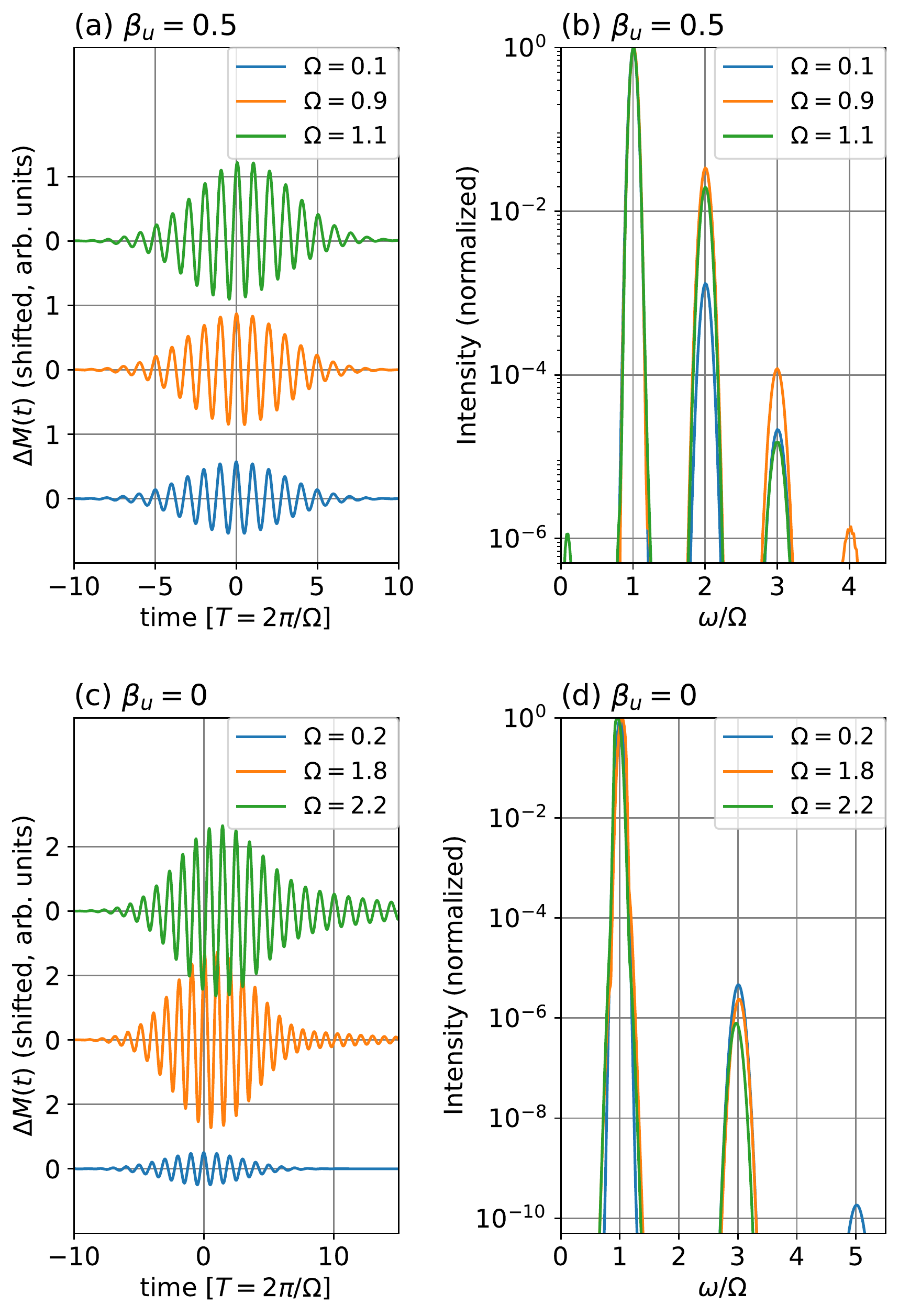}
\caption{
\hlr{
(left) Time profile of $\varDelta M(t)$ in the transverse-field Ising chain ($\aniso=1$) with
the ac Zeeman driving $\hextZnew(t)$ at frequencies well below, slightly below and above
the spin gap.
The static Zeeman field is $\beta_u=0.5$ (top) and $\beta_u=0$ (bottom),
where the spin gap is $\gapII=1$ and 2, respectively.
The other parameters are $\beta_0=0.1$ and $\gamma=0.1$.
(right) Corresponding power spectrum $I_M(\omega)$~\eqref{eq:powP}
plotted against $\omega/\Omega$
for $\beta_u=0.5$ (top) and $\beta_u=0$ (bottom).
}
}
\label{fig:spec_mag}
\efig

%%%%%%%%%%%%%%%%%%%%%%%%%%%%%%%%%%
%%%%%%%%%%%%%%%%%%%%%%%%%%%%%%%%%%
%%%%%%%%%%%%%%%%%%%%%%%%%%%%%%%%%%
%%%%%%%%%%%%%%%%%%%%%%%%%%%%%%%%%%
%############
\subsection{Numerical Results}
%#####
\bfig
\putfigraw{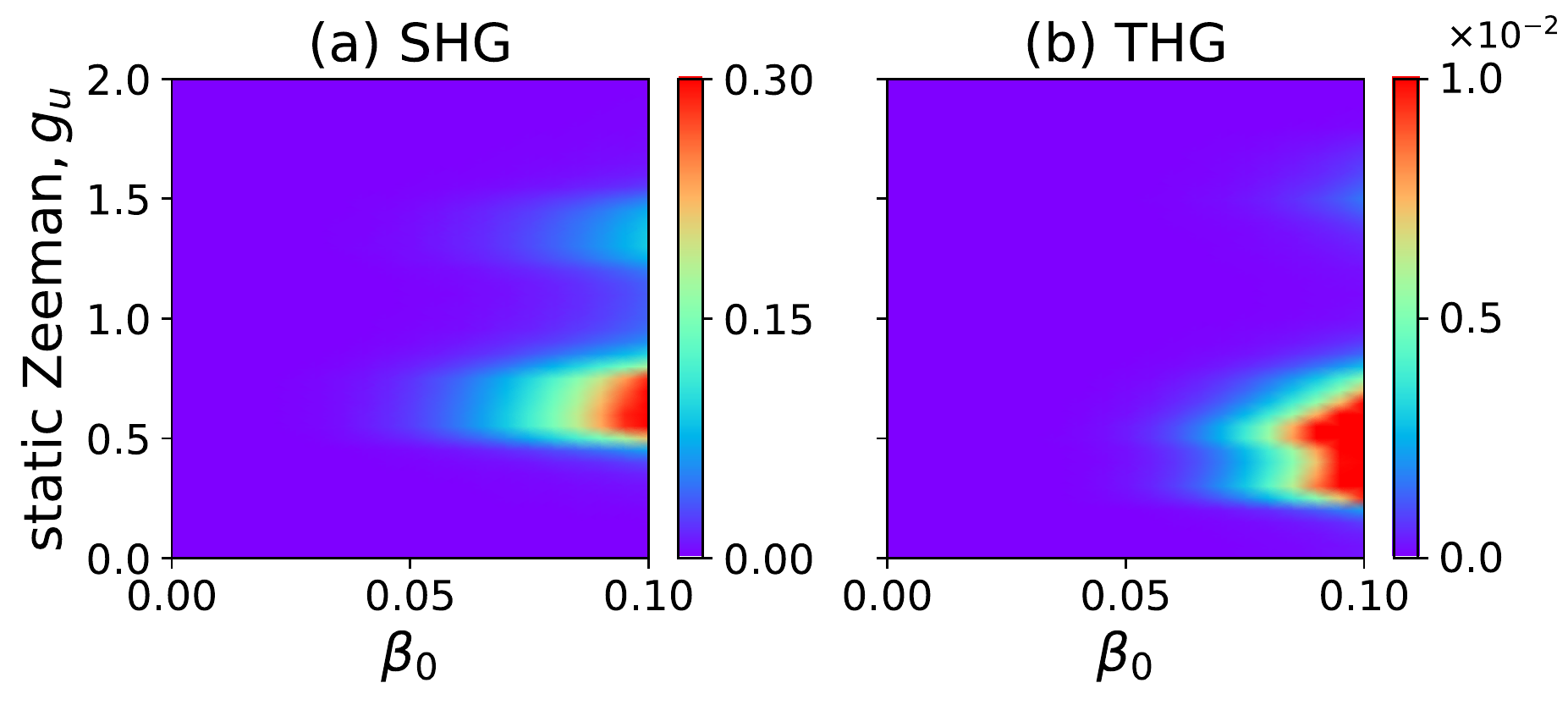}
\caption{
Intensities of SHG [$I_M(2\Omega)$] (left) and THG [$I_M(3\Omega)$] (right)
generated by ac-Zeeman-coupling driven magnetization dynamics 
in the transverse-field Ising chain ($\aniso=1$). 
The unit of intensity is chosen as $I_M(\Omega)$ at $\beta_0=0.05$
and $\tfig=0.5$.
The other parameters are set as $\gamma=0.1$ and $\Omega=0.5$.
}
\label{fig:color_mag}
\efig

Figure~\ref{fig:spec_mag}(a) shows a typical magnetization profiles 
in which we apply a laser pulse of Eq.~(\ref{eq:pulse_mag}) to the transverse-field Ising model. 
Here $\tfig=0.5$, i.e., $\gapII=1$,
and the driving frequencies are much smaller than or close to the gap.
Their normalized power spectrum $I_M(\omega)$ is shown in Fig.~\ref{fig:spec_mag}(b).
The even-order harmonics are present since $\tfig\neq0$
while we have confirmed that they are negligibly small in the absence of the static magnetic field,
\hlr{$\tfig=0$,
as shown in Fig.~\ref{fig:spec_mag}(d)}.
This is consistent with the symmetry argument in Sec.~\ref{sec:mag_model}.
Namely, we have numerically shown that the SHG is controllable
by the static magnetic field.

Whereas the SHG is strong both for $\Omega$ slightly above and below the spin gap,
the THG is stronger for $\Omega$ below the spin gap.
Similarly to the electric polarization and the spin current
discussed in the previous sections,
the harmonic peaks remain narrow even for $\Omega\sim\gapII$
because of relaxation.
Without relaxation, near-resonant driving causes strong real excitations,
which destroy the clear peak structures.
The interplay between the strong driving and relaxation
results in the strong and clear THG signals.

\begin{figure*}[t]\begin{center}
\includegraphics[width=16cm]{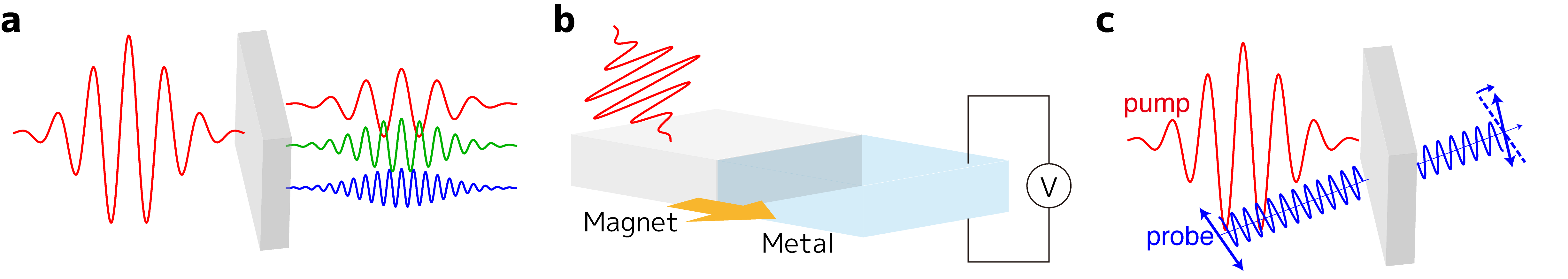}
\caption{
Schematic illustrations of experimental setups
to observe harmonic generation and spin currents.
(a) Strong THz laser induces nonlinear oscillations of
the electric polarization or the magnetization,
which gives rise to harmonic generation.
(b) Harmonic spin currents induced by strong THz waves in a magnet sample are injected into an attached metal, where the electromotive force is induced by the inverse spin Hall effect.
(c) An all-optical setup to detect harmonic spin currents
generated by strong GHz pumps.
The spin currents are detected by a weak probe high-frequency wave,
whose polarization changes according to the Faraday rotation.
}
\label{fig:experiments}
\end{center}\end{figure*}

We now systematically investigate the intensity of the HHG derived from magnetization dynamics.
Figure~\ref{fig:color_mag} shows $I_M(2\Omega)$ and $I_M(3\Omega)$ in the $(\beta_0,\tfig)$-plane
with $\Omega=0.5$, where the unit of intensity is taken as $I_M(\Omega)$ for $\beta_0=0.05$.
This value of $\Omega$ corresponds to $0.10$\,THz (0.52\,THz) for $J=10$\,K (50\,K).
Within the range of parameters in Fig.~\ref{fig:color_mag}, the SHG (THG) intensity tends
to monotonically increase with the ac Zeeman coupling $\beta_0$ and
becomes as large as 30\% (1\%) of our reference fundamental harmonic.

%#####
The SHG and THG show nonmonotonic behaviors in the static magnetic field $\tfig$.
These behaviors are understood by the multi-photon processes
in a perturbative viewpoint as follows.
As $\tfig$ increases from 0, the spin gap $\gapII$ decreases according to Eq.~\eqref{eq:gapII} 
(see also Fig.~\ref{fig:band_TFI}).
When the gap becomes smaller than 3$\Omega$ ($2\Omega$),
the three-(two-)photon process becomes significant and the THG (SHG) starts to increase.
One might expect that
the THG (SHG) enhances resonantly when
the resonance condition $3\Omega\,(2\Omega)=\gapII=2\epsilon_{\mathrm{II}}(k=\pi)$ is satisfied,
but this enhancement is not observed in Fig.~\ref{fig:color_mag}.
This is because the resonance occurs at $k=\pi$, but the ac Zeeman coupling at $k=\pi$
has no matrix elements between the upper and lower bands
[see Eqs.~\eqref{eq:HIImat} and \eqref{eq:hIImat}].
As $\tfig$ increases further to approach the quantum critical point $\tfig=1$ and the spin gap vanishes, 
the SHG and THG decrease. This is because near $\tfig=1$ the band dispersion $\epII$ becomes like a Dirac cone
and the density of state decreases.
As $\tfig$ increases from unity, the spin gap grows again and the SHG and THG increase. 
However, the HHG finally decreases to almost vanish 
when the gap becomes larger than $2\Omega$ and $3\Omega$ and the multi-photon processes are not significant.

%###
Figure~\ref{fig:color_mag} clearly shows that the SHG and THG are generally 
smaller in the forced ferromagnetic phase $(\tfig>1)$ than in the N\'{e}el one $(0<\tfig<1)$. 
This would be understood from the value of the magnetization along the $S^z$ direction. 
Namely, the initial magnetization $M_0$ becomes larger with increase of $\tfig$, 
and thus the application of the ac Zeeman field along the same direction 
leads to less-efficient magnetization oscillations.

%Finally, we
Let us estimate the required field amplitude
to observe the HHG through magnetization in our model.
We again follow the criterion discussed at the end of Sec.~\ref{sec:pol},
and regard $\beta_0=0.1$ as the required amplitude
from the $\Omega\sim1$ data in Fig.~\ref{fig:spec_mag}.
This pair of $\beta_0$ and $\Omega$ corresponds
to $B_0=0.74$\,T at 0.21\,THz for magnets of energy scale $J=10$\,K
and $B_0=3.7$\,T at 1.0\,THz for those of $J=50$\,K.
Compared to the HHG through the electric polarization
discussed in Sec.~\ref{sec:pol},
this amplitude is one-order more demanding.
This might be related to the difference between the origins of the SHG.
While we considered an inversion-asymmetric model in Sec.~\ref{sec:pol},
we here discuss another one with broken dynamical symmetry regarding
spin rotations.
The HHG by magnetization in inversion-asymmetric models
would merit future study.

%##############################
%##############################
%##############################
%##############################
%##############################
%##############################

\section{Experimental Protocols}\label{sec:experiments}
%######
We have shown that harmonic responses can be generated
in spin systems by THz and GHz electromagnetic waves.
In this section, we propose some ways to observe them,
and discuss how intense electromagnetic waves are required.

%#####
To detect the harmonic generation
through the electric polarization and the magnetization,
it is useful to observe the radiation from them by a spectrometer
as shown in Fig.~\ref{fig:experiments}(a). 
This is basically the same as that of HHG in semiconductors.

%#####
On top of detecting the radiation, 
the harmonic spin currents 
could be detected by the following two methods.
The first one is based on electric technology and attaching a spin-orbit-coupled metal 
on the sample magnetic insulator as shown in Fig.~\ref{fig:experiments}(b).
In this method, the generated ac spin currents are injected into the metal.
Then these spin currents are converted into the ac electric currents
through the inverse spin Hall effect~\cite{Saitoh2006,Kimura2007,Valenzuela2006}. 
Finally these electric currents are detected as the ac electric voltage
if the frequency is sufficiently low. As we already mentioned, high frequency electric voltage 
cannot be detected by the standard electric method and thus the frequency of the applied laser field 
should be equivalent to or smaller than several tens of THz~\cite{Wei2014}. 
The second one is an all-optical pump-probe method as shown in Fig.~\ref{fig:experiments}(c).
In this method, a weak high-frequency, e.g., visible light wave detects 
the magnetization dynamics driven by GHz or THz pump waves 
through the Kerr effect or the Faraday rotation.

%###############################
\section{Extremely Strong Fields}\label{sec:extreme}
\hlm{
We have been focusing mainly on the second and third harmonics
and the required field strengths to observe them.
We have shown that these harmonics could be observable with the state-of-the-art
intense lasers even though the laser-spin coupling is weak in principle.
Meanwhile, it is still of theoretical interest to investigate even higher-order harmonics
generated by extremely strong lasers that would not be available in the current technology.
In particular, since our spin models correspond to electron systems such as
semiconductors and superconductors, this investigation leads to
extending the correspondence of the models to that of the HHG spectra.
}

%######
%###################
\bfig
\putfigraw{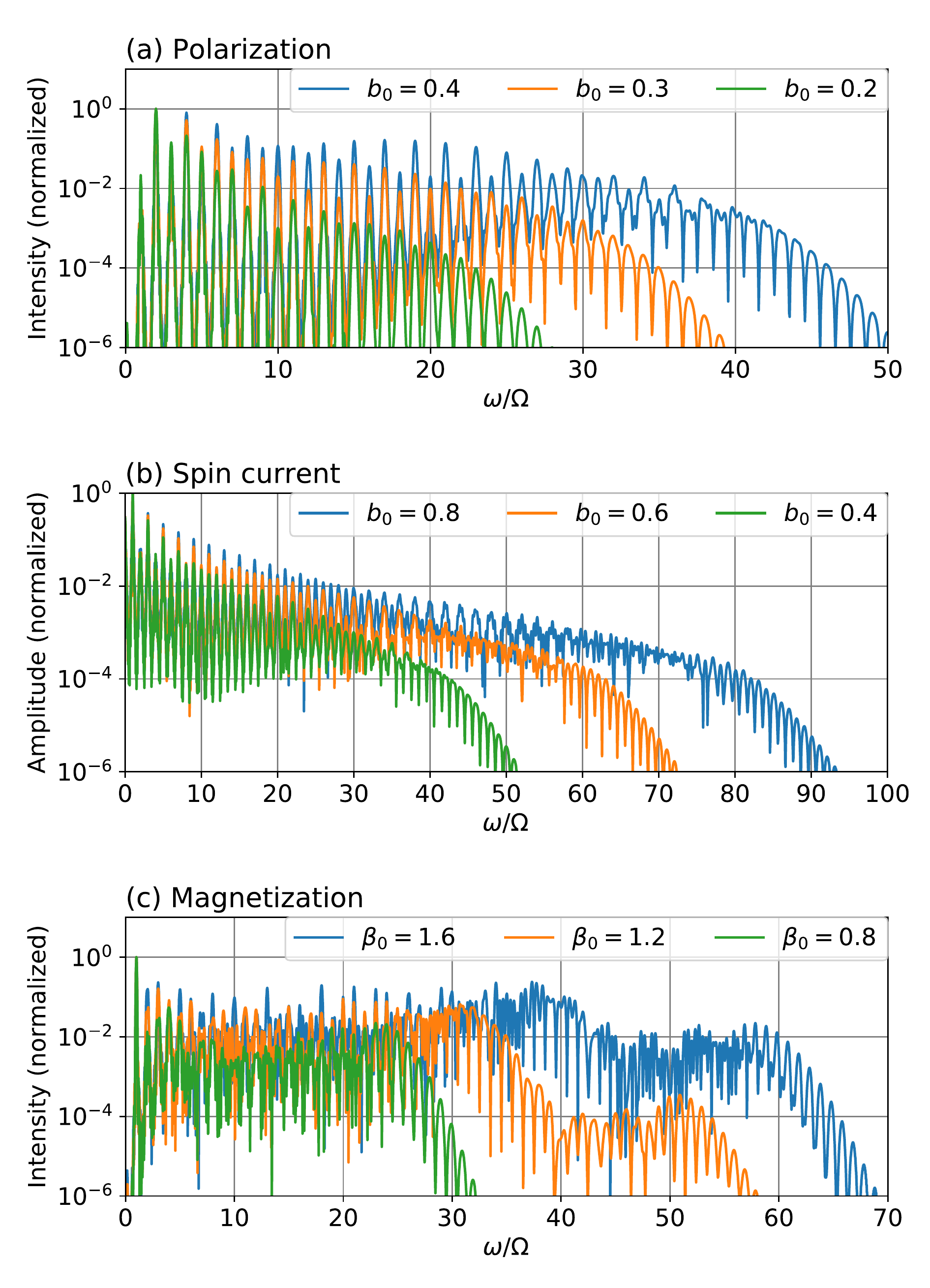}
\caption{
\hlm{
(a) Power spectrum $I_P(\omega)$~\eqref{eq:powP}
for extremely strong Zeeman drivings $\hextZ(t)$
with $b_0=0.2$, 0.3, and 0.4.
(b) Amplitude spectrum $|\Jspin(\omega)|$
for extremely strong Zeeman drivings $\hextZ(t)$
with $b_0=0.4$, 0.6, and 0.8.
In both (a) and (b),
the Hamiltonian is $\hfirst$ with
$\stagXY=0.1$ and $\stagZ=0.03$,
where the spin gap is $\gapI=0.21$,
and the driving frequency of $\hextZ(t)$ is well-below the gap $\Omega=0.02$.
Also, the data are normalized so that the highest peak
is unity for each spectrum.
(c) Power spectrum $I_M(\omega)$
in the transverse-field Ising chain ($\aniso=1$)
for extremely strong drivings with $\beta_0=0.8$, 1.2, and 1.6.
The parameters are $\beta_u=0.5$ where the spin gap is $\gapII=1.0$
and the driving frequency is well-below the gap $\Omega=0.1$.
The data are normalized so that the highest peak
is unity for each spectrum.
}
}
\label{fig:plateau}
\efig

\hlm{
Let us first consider
the HHG spectra by the polarization~\eqref{eq:powP}
in the model introduced in Sec.~\ref{sec:setup1}.
As we remarked in Sec.~\ref{sec:setup1},
our spin model in the fermion language is analogous to
the semiconductors.
Thus, if a very strong field is applied,
our spin system is expected to give HHG spectra similar to those of the semiconductors.
This is indeed the case as shown in Fig.~\ref{fig:plateau}(a),
in which we show the power spectrum $I_P(\omega)$~\eqref{eq:powP} for the Zeeman driving for example.
Note that the field amplitude $b_0$ is more than 10 times larger than those used in Sec.~\ref{sec:pol}.
We clearly see the plateau structure followed by
the rapid decrease of intensity.
The harmonic order at which the rapid decrease sets in
is known as the cutoff order in the semiconductor HHG~\cite{Ghimire2014},
which is roughly read from Fig.~\ref{fig:plateau}(a) as 20, 30, and 40
for $b_0=0.2$, 0.3, and 0.4, respectively.
This linear scaling of the cutoff order to the field amplitude is known as a unique feature of the semiconductor HHG.
Therefore, the similarity of our spin system
to the semiconductor system extends to that of HHG spectra
in these two systems.
}

%####
\hlm{
A similar correspondence to the semiconductor HHG is seen
in the spin-current spectrum $|\Jspin(\omega)|$ as well.
Figure~\ref{fig:plateau}(b) shows the results
for extremely strong Zeeman fields
with $b_0=0.4$, 0.6, and 0.8.
(see figure caption for the other parameters).
For, e.g., $b_0=0$,
we observe that the harmonic peak slowly decreases upto $\omega/\Omega \lesssim 80$
and then rapidly decays.
We interpret this as an approximate plateau with cutoff order about $80$.
The cutoff order thus defined is read out as $\sim60$ and $\sim40$ for $b_0=0.6$ and $0.4$, respectively.
Thus the cutoff order roughly scales linearly with $b_0$
in line with the semiconductor HHG.
%As remarked in the previous section,
%this is of theoretical interest in consolidating the correspondence
%between our spin model and semiconductors.
}

%####
\hlm{
Finally we investigate the HHG spectra by the magnetization~\eqref{eq:powM}
for the transverse-field Ising chain ($\aniso=1$).
As remarked in Sec.~\ref{sec:mag},
this model is mapped to a BCS-type model
of superconductors.
Figure~\ref{fig:plateau}(c) shows $I_M(\omega)$
for $\beta_0=0.8$, 1.2, and 1.6 that are
roughly 10 times larger than those we have considered in Sec.~\ref{sec:mag}.
For $\beta_0=0.8$ we observe a plateau-like behavior upto $\omega/\Omega\lesssim25$
and then a rapid decrease.
The width of the plateau-like behavior increases almost linearly, roughly speaking,
with the field amplitude $\beta_0$.
It is more remarkable that the second plateau emerges for higher fields
and observed for $\beta_0=1.2$ and $1.6$.
The second plateau has not been seen in Figs.~\ref{fig:plateau}(a) and (b), for which the Hamiltonian is analogous to semiconductors.
Thus the second plateau might possibly related to superconductors, to which
the present model is analogous.
We leave further study on this relation as a future work.
}

%##############################
%##############################
%##############################
%##############################
\section{Conclusions}\label{sec:conclusions}
We have investigated the harmonic generation and harmonic spin currents in magnetic insulators. 
To this end, we have considered simple but realistic models of quantum spin chains and 
studied the laser-driven nonlinear dynamics by means of the quantum master equation. 
In Sec.~\ref{sec:setup1}, we have introduced the inversion-asymmetric spin chain
to study the HHG by electric polarization and spin current. 
Through the Jordan-Wigner transformation, 
the model is exactly mapped to a two-band fermion model like semiconductors. 
We have confirmed that both intra- and inter-band transitions of the Jordan-Wigner fermions 
are relevant to generate harmonic peaks similarly to the HHG of semiconductors.
On the other hand, we have focused on the transverse-field Ising chain 
to explore HHG by magnetization dynamics in Sec.~\ref{sec:mag}, 
and the chain is mapped to a fermion model with a BCS-type Hamiltonian. 
Calculating the quantum dynamics under pulse lasers and relaxation in Secs.~\ref{sec:pol}-\ref{sec:mag},
we have shown that the harmonic peaks can appear in the electric polarization,
the spin current, and the magnetization.
As shown in Figs.~\ref{fig:spec_pol_Z}, \ref{fig:spec_sc_Z}, and \ref{fig:spec_mag},
these harmonic peaks have been obtained in a well-defined manner
thanks to the relaxation taken in our quantum master equation.
\hlm{For hypothetical strong fields,
we have obtained the harmonic spectra involving plateaus
in our spin models
and pointed out the correspondence to the spectra in semiconductors
and superconductors
(see Fig.~\ref{fig:plateau}).}

\hlm{For realistic field strength within the state-of-the-art technology,
the obtained harmonics become large enough to be experimentally observed.}
The required ac electirc-field strength is typically $E_0=100$\,kV/cm--1\,MV/cm. 
The THz and GHz waves with these field strengths could be achieved within the current laser 
technology~\cite{Hirori2011,Dhillon2017,Liu2017,Mukai2016}. 
The data in Tables I-III would be useful to semi-quantitatively estimate the required laser field 
to experimentally create HHG in magnets. 
We have shown that the harmonic peaks are not sensitive
to the driving frequency and the field strength is more important
for a successful detection.
It is noteworthy that the SHG from magnetization dynamics has been shown controllable
by static Zeeman fields.
This controllability applies to a wide class of magnetic insulators.

We have proposed some experimental ways of observing HHG in magnetic insulators in Sec.~\ref{sec:experiments}. 
In addition to the optical method, we have considered electric ways of detecting high-harmonic spin currents 
(see Fig.~\ref{fig:experiments}).   
An intense GHz wave (i.e., sufficiently low-frequency laser or electromagnetic wave) 
is necessary to use the electric methods.

As we mentioned in Introduction, the HHG is a typical and simple nonlinear optical phenomenon in solids. 
Therefore, our estimate for the required ac field strength ($E_0=100$\,kV/cm--1\,MV/cm) 
would serve as a reference value not only for future HHG experiments in real magnets
but also for other nonlinear magneto-optical effects such as Floquet engineering of magnetism~\cite{Sato2014,Sato2016}, 
dc spin current rectification with THz or GHz waves~\cite{Ishizuka2019a,Ishizuka2019b}, inverse Faraday effects~\cite{Takayoshi2014a,Takayoshi2014b}, etc.  

%######
Extensions of the present work to magnonic systems are future work of interest.
A difference from our present model mappable to the Jordan-Wigner fermions
is that the magnons may exhibit resonances to the external field
and stronger signals could be obtained in certain conditions.
Of course, more quantitative theoretical evaluations specific to each material
become important when one interprets concrete experiments.

%%%%%%%%%%%%%%%%%%%%%%%%%%%%%%%%%%%%%%
%%%%%%%%%%%%%%%%%%%%%%%%%%%%%%%%%%%%%%
%%%%%%%%%%%%%%%%%%%%%%%%%%%%%%%%%%%%%%
%%%%%%%%%%%%%%%%%%%%%%%%%%%%%%%%%%%%%%
\section*{Acknowledgements}
Fruitful discussions with Hiroaki Ishizuka, Daichi Hirobe, and Masashi Fujisawa
are gratefully acknowledged.
T.N.I. is supported by JSPS KAKENHI Grant No.~JP18K13495.
M.S. is supported by JSPS KAKENHI Grant No. 17K05513 and 
a Grant-in-Aid for Scientific Research on Innovative Areas ``Quantum Liquid Crystals''  (Grant No. JP19H05825).

%################################
\appendix
\section{HHG selection rule for magnetization dynamics}\label{app:ds}
\hl{
We supplement the argument in Sec.~\ref{sec:mag_model},
where the even-order harmonics are shown to vanish
when the static Zeeman field is absent $\beta_u=0$.
The key equation is
\begin{align}
	M(t) = -M(t+T/2),\label{eq:app:aim}
\end{align}
which is satisfied by the time-dependent expectation value of
the total magnetization.
In this appendix, we derive the above equation
by more careful calculations.
}

\hl{
For simplicity, we restrict ourselves to the Ising case $\aniso=1$.
%in which the ground state is the N\'{e}el state along the $S^x$ axis.
In this case, both the static Hamiltonian $\hxy$ and the ground state
are invariant under the global $\pi$ rotation around the $S^x$ axis $\pix$,
$\pix\hS^\alpha_j\pixd=-\hS^\alpha$ $(\alpha=y,z)$ and $\pix\hS^x_j\pixd=\hS^x$.
Note that our argument can easily be generalized to the general case $\epsilon\neq1$, 
where the symmetry axis depends on $\epsilon$.
}

\hl{Let us express the symmetry in terms of the Jordan-Wigner
fermions. As one can check easily, The global rotation $\pix$
leads to
%\begin{align}
	$\pix \ann_j \pixd = (-1)^j \cre_j$,
%\end{align}
which implies, in the Fourier transform, that
%\begin{align}
	$\pix \hat{d}_k \pixd  = \hat{d}_{\pi-k}^\dag$.
%\end{align}
Then the rotational invariance of $\hxy$ is translated
into the $2\times2$-matrix representation~\eqref{eq:HIImat} as
\begin{align}
	\hsecondmat(\pi-k) = \sigma_y \hsecondmat(k)\sigma_y\label{eq:H2symm}
\end{align}
for $\beta_u=0$.
This means that there is one-to-one correspondence
between the energy eigenstates of $\hsecondmat(\pi-k)$ and $\hsecondmat(k)$,
which leads to
\begin{align}
	\rho(\pi-k,\tini) = \sigma_y \rho(k,\tini)\sigma_y.\label{eq:app:init}
\end{align}
at the initial time.
}

\hlr{Next, we consider the dynamical symmetry,
supposing the multi-cycle limit, at which
$T/\tfwhm\gg1$ and $\beta(t)$ is approximately sinusoidal.
In this situation, the ac Zeeman field satisfies $\beta(t+T/2)=-\beta(t)$,
and the total Hamiltonian $\hat{H}(t)=\hsecond+\hextZnew(t)$
has the dynamical symmetry $\hat{H}(t) = \pix\hat{H}(t+T/2)\pixd$.
In the fermion language, this dynamical symmetry reads
\begin{align}
	H(\pi-k,t) = \sigma_y H(k,t+T/2)\sigma_y.\label{eq:app1}
\end{align}
We remark that the Lindblad operators satisfy similar relations
\begin{align}
L_{\pi-k}=\ee^{\ii \varphi_k}\sigma_yL_{k}\sigma_y,	\label{eq:app2}
\end{align}
where $\varphi_k$ is some real number.
}

\hl{
Equations~\eqref{eq:app1} and \eqref{eq:app2}
relate the dynamics of $\rho(k,t)$ and $\rho(\pi-k,t)$.
To see this, we symbolically represent
the quantum master equation~\eqref{eq:QME}
by using the Liouvillian super operator
\begin{align}
	\frac{\dd}{\dd t} \rho(k,t) = \mathcal{L}_k(t) \rho(k,t).\label{eq:Liouville1}
\end{align}
Equations~\eqref{eq:app1} and \eqref{eq:app2}
lead to
\begin{align}
	\frac{\dd}{\dd t} \rho(\pi-k,t)
	&= \mathcal{L}_{\pi-k}(t) \rho(\pi-k,t)\notag\\
	&=\sigma_y\mathcal{L}_{k}(t+T/2)\sigma_y \rho(\pi-k,t),
\end{align}
and thus to
\begin{align}
	\frac{\dd}{\dd t} \sigma_y\rho(\pi-k,t)\sigma_y
	=\mathcal{L}_{k}(t+T/2)\sigma_y \rho(\pi-k,t)\sigma_y.\label{eq:Liouville2}
\end{align}
Comparing Eqs.~\eqref{eq:Liouville1} and \eqref{eq:Liouville2},
we obtain
\begin{align}
	\sigma_y\rho(\pi-k,t)\sigma_y = \rho(k,t+T/2)\label{eq:app:evolsymm}
\end{align}
if $\sigma_y\rho(\pi-k,\tini)\sigma_y=\rho(k,\tini+T/2)$.
This condition is actually satisfied because of the following reasons.
First, Eq.~\eqref{eq:app:init} holds true.
Second, the time evolution of $\rho(k,t)$
from $t=\tini$ to $\tini+T/2$ is negligible
because we have taken such a small $\tini$
that the ac Zeeman field is negligible at $t\sim\tini$.
}

\hl{
Finally, we discuss the magnetization,
which is given in the $2\times2$-matrix representation by
\begin{align}
	M_k = -\sigma_z
\end{align}
independently of $k$.
Thus, Eq.~\eqref{eq:app:evolsymm} leads to
the following relations for the expectation values
of magnetization:
\begin{align}
	M(\pi-k,t) &= \text{tr} [(-\sigma_z)\rho(\pi-k,t)]\notag\\
	&= \text{tr}[(-\sigma_z) \sigma_y \rho(k,t+T/2)\sigma_y]\notag\\
	&= -\text{tr}[(-\sigma_z) \rho(k,t+T/2)]\notag\\
	&= -M(k,t+T/2),
\end{align}
where we have used the cyclic property of the trace
and $\sigma_y\sigma_z\sigma_y=-\sigma_z$.
Therefore, the total magnetization $M(t)$ satisfies
\begin{align}
	M(t) &= \sum_k M(k,t)\notag\\
		&= \sum_k M(\pi-k,t+T/2)\notag\\
		&= -M(t+T/2).
\end{align}
Thus we have obtained Eq.~\eqref{eq:app:aim}.
}

\end{document}